\documentclass[%
 reprint,
superscriptaddress,
showpacs, 
nofootinbib,
nobibnotes,
amsmath,amssymb,
aps,
pra,
]{revtex4-1}

\usepackage{graphicx}
\usepackage{dcolumn}
\usepackage{bm}
\usepackage{amsmath}
\usepackage{amsfonts}
\usepackage{amssymb}

\begin{document}

\preprint{APS/123-QED}
 
\title{Analysis of terahertz generation using tilted-pulse-fronts}

\author{Koustuban Ravi}
\email {koust@mit.edu}
\affiliation{
Center for Free-Electron Laser Science, DESY,Notkestra$\beta$e 85, Hamburg 22607, Germany
}
\affiliation{
Research Laboratory of Electronics, Massachusetts Institute of Technology, 50 Vassar Street, Cambridge MA 02139.
}

\author{Franz X. K\"artner}
\affiliation{
Center for Free-Electron Laser Science, DESY,Notkestra$\beta$e 85, Hamburg 22607, Germany
}%
\affiliation{
Research Laboratory of Electronics, Massachusetts Institute of Technology, 50 Vassar Street, Cambridge MA 02139.
}
\affiliation{
Department of Physics, University of Hamburg, Hamburg 22761, Germany 
}%

\begin{abstract}
A 2-D spatio-temporal analysis of terahertz generation by optical rectification of tilted-pulse-fronts is presented. Closed form expressions of terahertz transients and spectra in two spatial dimensions are furnished in the undepleted limit. Importantly, the analysis incorporates spatio-temporal distortions of the optical pump pulse such as angular dispersion, group-velocity dispersion due to angular dispersion, spatial and temporal chirp as well as beam curvature. The importance of the radius of curvature to the tilt-angle and group-velocity dispersion due to angular dispersion to terahertz frequency, conversion efficiency and peak field is revealed.In particular, the deterioration of terahertz frequency, efficiency and field at large pump bandwidths and beam sizes by group velocity dispersion due to angular dispersion is analytically shown. 
\end{abstract}

\maketitle

\section{Introduction}
The generation of terahertz radiation, i.e. electromagnetic radiation in the frequency range spanning $0.1-10$ THz has experienced a recent surge in interest. The large peak fields that can be generated at these frequencies in combination with long wavelengths offers unique opportunities to manipulate the motion of charged particles. As a result, a number of applications such as compact charged particle acceleration \citep{nanni2015,arya2016,palfalvi2014} and streaking \citep{zhang2018,kealhofer2016,ronnyhuang16}, control of emission from nanotips \citep{ropers2016} and higher-harmonic generation in solids \citep{schubert2014} have emerged. In addition, the proximity of these frequencies to lattice vibrations makes them uniquely amenable to probing a number of fundamental scientific phenomena \citep{kampfrath2013}.

Among various methods of terahertz generation \cite{gallerano2004,gold1997}, laser-driven nonlinear optical methods have gained ground owing to improvements in solid-state laser technology \citep{zapata2016} as well as the intrinsic synchronization they offer-which is valuable to scientific investigations. Furthermore, the long assumed efficiency limitation of this approach due to the large disparity between optical and terahertz photons has been dispelled by experimental demonstrations of cascaded difference frequency generation/optical rectification \cite{huang2015}. Specifically, this class of approaches enables the repeated energy down conversion of optical pump photons to yield energy conversion efficiencies at the $\eta=1\%$ level \citep{vicario14}. 

Of various nonlinear optical approaches to generate terahertz radiation, optical rectification of angularly dispersed beams or tilted-pulse-fronts \citep{hebling02} in lithium niobate has become ubiquitous. The approach is accessible by commercially available Titanium Sapphire lasers at 800 nm and has yielded very high conversion efficiencies and pulse energies to date \citep{fulop2011}.

However, the complex spatio-temporal shaping of the pump pulse introduces many subtleties to the physics governing the approach. The initial treatment of the problem by 1-D spatial models \citep{jewariya2009,fulop2011,fulop10} with effective parameters, while informative, do not account for the intrinsic non-collinearity of the problem.Numerical models considering multiple spatial dimensions \cite{zhong2015} as well as the effects of pump depletion \citep{ravi15} furnish accurate quantitative predictions but do not provide the intuition and understanding that arises from analytic approaches. Initial analytic or semi-analytic 2-D models were  previously reported \citep{shuvaev2007,bakunov2008,bakunov2011,bakunov2014}. 

Here, we further develop analytic methods which accurately account for the various spatio-temporal distortions that accompany pulses with tilt-pulse-fronts. These include the effects of group velocity dispersion due to angular dispersion (GVD-AD), spatial and temporal chirp as well as effects of beam curvature. 

These models in the undepleted limit straddle the middle ground between 1-D models and 2-D numerical models incorporating pump depletion. The analysis sheds light on various spatio-temporal aspects of terahertz generation by tilted-pulse-fronts. Closed form expressions of terahertz spectra in transverse momentum, spatial and temporal domains are furnished.

 The analysis reveals the importance of group-velocity dispersion due to angular dispersion. In particular, it reduces the terahertz frequency, conversion efficiency and peak field at large bandwidths and beam sizes. This has ramifications on energy scaling and use of short pump pulses. While this phenomenon has been partially revealed through experiments and numerical studies, here this is shown via analytic approaches and closed-form expressions. The presented expressions are quantitatively useful for room temperature terahertz generation in lithium niobate while they provide qualitative insights at cryogenic temperatures. In particular,this undepleted analysis underscores the importance of considering cascading effects or depletion effects in the low absorption limit. 
 
In Section II, we present the theoretical formulation. In Section III, detailed calculations of terahertz spectra, conversion efficiency and peak field are furnished. We conclude in Section IV. Appendices are provided for self-consistency.

\section{Theory}
\begin{figure}
\centering
\includegraphics[scale=0.4]{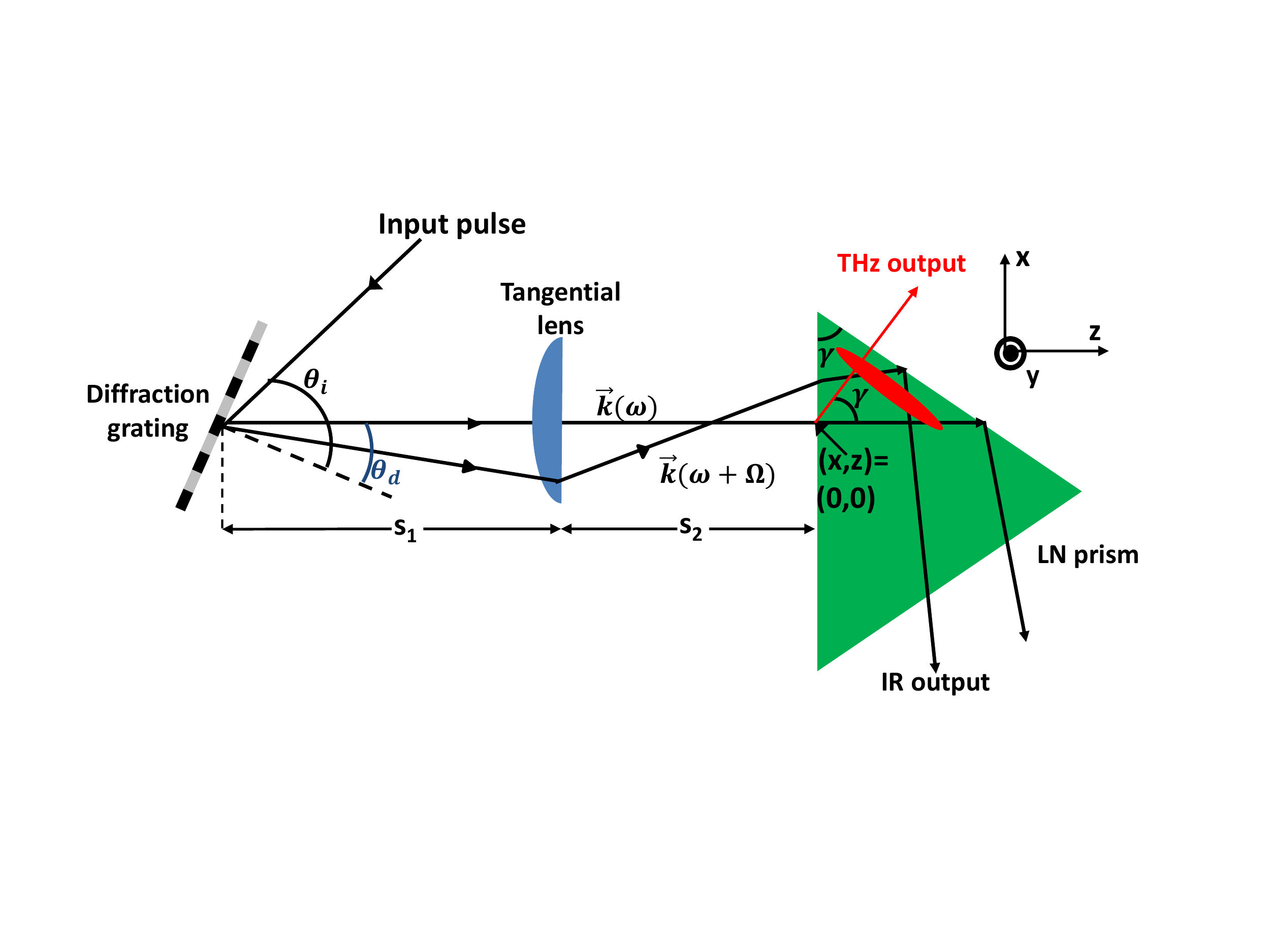}
\caption{A tilted-pulse-front setup comprising of a diffraction grating and an imaging system. The angularly dispersed pulse produces a tilted-pulse-front (red ellipse) with tilt-angle $\gamma$, resulting in terahertz radiation propagating at an angle $\gamma$ with respect to the direction of pump pulse propagation.}
\label{fig1}
\end{figure}

\subsection{General approach}

In the most general case, the generation of terahertz radiation by tilted-pulse-fronts follows from a solution of the coupled nonlinear wave equations for terahertz and optical waves \cite{ravi15}. However, since the approach outlined in this paper is analytic, we restrict ourselves to an undepleted solution. A typical tilted-pulse-front setup is depicted in Fig.\ref{fig1}. It consists of a diffraction grating off which the input pump pulse is scattered and then imaged into a crystal. As shown in Fig.\ref{fig1}, various wave vectors $\vec{k}(\omega)$ of optical components at corresponding angular frequencies $\omega$ are angularly dispersed in the $x-z$ plane, producing an intensity front which is tilted with respect to the propagation direction by an angle $\gamma$. 

The angular dispersion imparted to the pump pulse produces spatio-temporal coupling effects only in one plane, hence making a two-dimensional $(x,z)$ spatial model sufficient to capture the essential physics of the system. Various $x-z$ slices in the $y$ direction, are replicas weighted by the intensity profile of the pump pulse, which can be accounted for by appropriate scaling factors. 

The scalar wave equation governing terahertz spectral components at angular frequencies $\Omega\geq 0$ delineated by the phasor $E_{THz}(\Omega,x,z)$ is given by Eq.\ref{wav_eq1} (For additional details on transformation between phasor quantities and real fields, the reader is referred to the Appendix). The first term on the right hand side (RHS) of Eq.\ref{wav_eq1} corresponds to terahertz absorption while the second term on the RHS of Eq.\ref{wav_eq1} delineates the nonlinear polarization term $P_{THz}$ driving the generation of terahertz radiation at angular frequency $\Omega$.  Essentially, $P_{THz}$ is an ensemble of difference frequency generation processes proportional to the second order nonlinear susceptibility $\chi^{(2)}$ between various spectral components of the optical pump $E_{op}(\omega,x,z)$ and is given by Eq.\ref{p_thz}. 

\begin{subequations} 
\begin{gather}
\nabla^2 E_{THz}(\Omega,x,z) +k^2(\Omega)E_{THz}(\Omega,x,z) =\nonumber\\
jk(\Omega)\alpha(\Omega) E_{THz} (\Omega,x,z) -\frac{\Omega^2}{\varepsilon_0c^2}P_{THz}(\Omega,x,z)\label{wav_eq1}\\
P_{THz}(\Omega,x,z)  =  \varepsilon_0\chi^{(2)}\int_{0}^{\infty}E_{op}(\omega+\Omega,x,z)E_{op}^{*}(\omega,x,z)d\omega\label{p_thz}
\end{gather}
\end{subequations} 

Performing a Fourier decomposition on Eq.\ref{wav_eq1}, we obtain the ordinary differential equation with respect to $z$ for $E_{THz}(\omega,k_x,z)$ in Eq.\ref{wave_eq_FT}. Here, $k_x$ corresponds to the transverse momentum in the $x$ direction. 
 
\begin{gather} 
\frac{\partial^2 E_{THz}(\Omega,k_x,z)}{\partial z^2} +k_z^2(\Omega,k_x) E_{THz}(\Omega,k_x,z) = \nonumber\\
jk(\Omega)\alpha(\Omega) E_{THz}(\Omega,k_x,z) -\frac{\Omega^2 P_{THz}(\Omega,k_x,z)e^{-j\Omega v_g^{-1}z}}{\varepsilon_0 c^2} \label{wave_eq_FT}
\end{gather}

To solve the above, we set $E_{THz}(\Omega,k_x,z) = A_{THz}(\Omega,k_x,z)e^{-jk_z(\Omega,k_x)z}$, where $k_z(\Omega,k_x)=\sqrt{k(\Omega)^2-k_x^2}$ is the z-component of the terahertz wave vector $k(\Omega)$.In using the above ansatz , a $1/\text{cos}\gamma$ pre-factor is obtained for the absorption term due to the pre-factor $k(\Omega)/k_z(\Omega,k_x)\approx 1/\text{cos}\gamma$. Here, $\gamma$ is the angle at which the generated terahertz wave propagates with respect to the pump beam and is also equal to the tilt-angle. Furthermore, the initial condition $A(\Omega,k_x,z=0)=0$ is assumed to delineate the absence of any terahertz field at the entrance boundary of the crystal. The general solution for $E_{THz}(\Omega,k_x,z)$ is subsequently obtained as follows : 
\begin{subequations}
\begin{gather}
E_{THz}(\Omega,k_x,z) =\nonumber\\
e^{-\frac{\alpha(\Omega)z}{2\text{cos}\gamma}}\int_{0}^{z} \frac{-j\Omega^2}{2\varepsilon_0 k_z c^2}P_{THz}(\Omega,k_x,z')e^{j\Delta k(\Omega,k_x)z' +\frac{\alpha(\Omega) z'}{2\text{cos}\gamma}}dz'\label{e_thz_kx_a}\\
\Delta k(\Omega,k_x) = k_z-\Omega n_gc^{-1}\label{e_thz_kx_b}
\end{gather}
\end{subequations}

In Eq.\ref{e_thz_kx_b}, $\Delta k(\Omega,k_x)$ represents phase-mismatch between the pump and terahertz fields. Notice that setting Eq.\ref{e_thz_kx_b} to zero,yields the well-known relationship for the tilt angle $\gamma = \text{cos}^{-1}(n_g/n_{THz})$, where $n_g$ and $n_{THz}$ are the optical group and terahertz phase refractive indices respectively.
 
\subsection{Nonlinear polarization}

Having laid out the general framework Eq.(\ref{wav_eq1}), we proceed to obtain solutions for the generated terahertz field for an optical pump pulse with spatio-temporal distortions. We proceed by defining the optical pump spectrum $E_{op}(\omega,x,z)$ in Eq.\ref{eq:TPF_fielda}. Due to the large number of variables, we provide a glossary in Table.\ref{var_list} in the appendix for quick reference.

\begin{widetext}
\begin{gather}
E_{op}(\omega,x,z) = E_0 e^{-\frac{\Delta\omega^2\tau_0^2}{4}}e^{-\frac{(x-\zeta(z)\Delta\omega)^2}{w_0^2}}e^{-j\phi_0\Delta\omega^2}
e^{-j\frac{\omega n(\omega)(x-\zeta(z)\Delta\omega)^2}{2cR_0(z)}}
e^{-j[k_0+v_g^{-1}\Delta\omega+k_m\Delta\omega^2]z}e^{-jk_0\beta_0\Delta\omega x}e^{-jk_T\Delta\omega^2x}\label{eq:TPF_fielda}
\end{gather}
\end{widetext}

Equation \ref{eq:TPF_fielda} delineates the spectral components of the optical field at angular frequencies $\Delta\omega=\omega-\omega_0$ (where $\omega_0$ is the center frequency of the pump) with various spatio-temporal distortions. The first exponential term describes a spectrum with transform limited (TL) pulse duration $\tau_0$. The second exponential term on the RHS of Eq.\ref{eq:TPF_fielda}, represents a spatially chirped transverse beam with $e^{-2}$ radius $w_0$. Here $\zeta(z)= dx_c(z)/d\omega$ and delineates that different spectral components are centered at different spatial locations $x_c(z)$. The $z$ dependence of $\zeta,x_c$ arises from angular dispersion in the pump beam, which causes their spectral positions to change. The third and fourth exponential terms represent temporal chirp with group delay dispersion (GDD) $\phi_0$  that may be imparted to the beam and phases accrued due to the finite radius of curvature $R_0(z)$ of the beam respectively. The fifth exponential term represents the $z$-directed momentum of the optical field. As shown, this is represented by a polynomial expression about the wave number $k_0=\omega_0n(\omega_0)/c$ at the central angular frequency $\omega_0$, accounting for the group velocity $v_g$ and material dispersion via the parameter $k_m$. Here $n(\omega_0)$ represents the refractive index of material at the central angular frequency $\omega_0$. The penultimate and final terms are the most important terms for a beam forming a tilted-pulse-front (TPF). These correspond to the angular dispersion term $\beta_0=d\theta/d\omega$ and the group-velocity dispersion due to angular dispersion (GVD-AD) term denoted by $k_T$. The latter represents the fact that the angles of various spectral components are not distributed equally for equal increments in frequency $\Delta\omega$.

Using, the above expression for the optical spectrum $E_{op}(\omega,x,z)$ in the nonlinear polarization term defined by Eq.\ref{p_thz}, one obtains the following expression for $P_{THz}(\Omega,x,z)$ in Eq.\ref{pol_thz_ST2}.

\begin{widetext}
\begin{subequations}
\begin{gather}
P_{THz}(\Omega,x,z) = \frac{\sqrt{2\pi}}{\tau}\bigg[\varepsilon_0E_0^2\chi^{(2)}e^{-\frac{\Omega^2\tau^2}{8}}e^{-2\frac{[\phi_0+k_Tx +k_mz]^2\Omega^2}{\tau^2}}e^{-\frac{2x^2}{w_0^2}}\bigg]
e^{-j\bigg[\frac{n_g}{2cR_0(z)}+\frac{8\zeta(z) k_T}{\tau^2w_0^2}\bigg]\Omega x^2}\nonumber\\
\times e^{-jv_g^{-1}\Omega z}e^{-j\bigg(k_0\beta_0+\frac{2\zeta(z)\phi_0}{\tau_0^2w_0^2/4+\zeta(z)^2}-\frac{\omega_0n(\omega_0)\zeta(z)}{cR_0(z)}\bigg)\Omega x}\label{pol_thz_ST2}\\
\tau = \bigg(\tau_0^2+\frac{4\zeta(z)^2}{w_0^2}\bigg)^{\frac{1}{2}} \label{pol_thz_ST3}
\end{gather}
\end{subequations}
\end{widetext}

Equation \ref{pol_thz_ST2} exhibits a number of spatio-temporal coupling effects. The first exponential term in Eq.\ref{pol_thz_ST2} shows how the terahertz pulse duration is given by $\tau/\sqrt{2}$ for$k_T,k_m,\phi_0=0$. The second exponential term depicts an increase in terahertz pulse duration due to the temporal chirp $\phi_0$ as well as the GVD-AD ($k_T$) and GVD-MD terms ($k_m$). Specifically, while GVD-AD causes an increase in the pulse duration in the transverse ($x$) direction, the duration in the propagation direction $z$ increases due to material dispersion.

A further illustration of spatio-temporal coupling is evident by the change in effective pulse duration from $\tau_0$ to $\tau$ in Eq.\ref{pol_thz_ST3}. Here, the spatial-chirp $\zeta$ is seen to increase the effective duration. In a spatially chirped beam, all spectral components do not overlap in space. therefore, only a certain fraction of the original bandwidth is overlapped to produce terahertz radiation and this manifests via an increased effective pulse duration. 

The third exponential term simply shows that the beam size of the generated terahertz radiation would correspond to a beam radius $w_0/\sqrt{2}$, as with any second order nonlinear process. 

\begin{figure}
\centering
\includegraphics[scale=0.275]{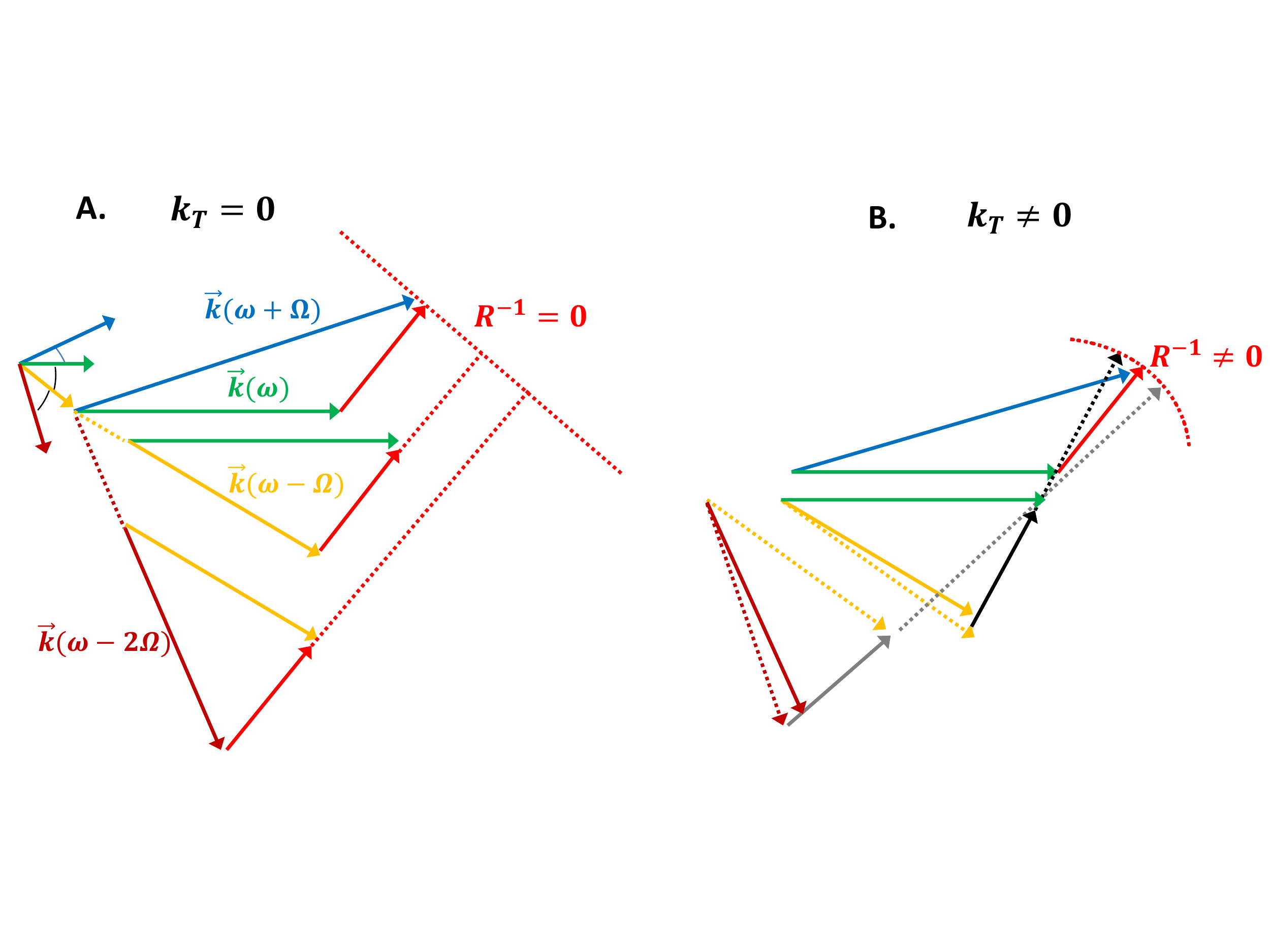}
\caption{Explaining the finite radius of curvature produced by a finite value of group-velocity dispersion due to angular dispersion (GVD-AD) $k_T$. (a) $k_T=0$ results in the wave vectors of various pump frequency components being spaced by roughly equal angles, which results in all terahertz vectors produced by each pair $\vec{k}(\omega+m\Omega),\vec{k}(\omega+(m-1)\Omega)$ being parallel to each other. This results in a flat terahertz phase-front or $R^{-1}=0$. (b) When $|k_T|>0$, different pump frequency components (dotted lines) are displaced from their original angles(solid lines), producing terahertz vectors pointing in slightly different directions. This results in a curved phase-front.}
\label{fig2}
\end{figure}

The fourth exponential term shows that the polarization contains a radius of curvature which is different from that of the incident pulse. The schematic in Fig.\ref{fig2}, depicts why this is the case. In the presence of a finite $k_T$, the phase-matched terahertz directions are not all the same and thus produce a phase-front with some curvature. Since, the extent of this spread in terahertz directions shall be proportional to the incident pump bandwidth, the situation is more adverse for shorter pump pulse durations. This is yet another example of a spatio-temporal coupling effect that arises in TPFs that can only be accounted for via an appropriate modelling of the optical pump field. Therefore, the above equations already indicate the great importance of the $k_T$ term in determining the properties of the generated terahertz radiation. While prior work has illustrated the importance of GVD-AD numerically \cite{ravi2014}, here we are able to provide an insight into its effects analytically.

\subsection{Tilt angle}

A critical aspect of Eq.\ref{pol_thz_ST2} is delinated in the two final exponential phase terms, which represent a line of constant phase $z+\text{tan}\gamma x=0$. Here, $\gamma$ is the pulse-front-tilt angle and is given by Eq.\ref{TPF:angle}.

\begin{gather}
\text{tan}\gamma = k_0\beta_0 v_g  +\frac{2\zeta(z)\phi_0 v_g}{\frac{\tau_0^2w_0^2}{4}+\zeta(z)^2}-\frac{\omega_0n(\omega_0)\zeta(z) v_g}{cR_0(z)}\label{TPF:angle}
\end{gather}

Each term in Eq.\ref{TPF:angle} represents a different source of pulse-front-tilt.  The first term in Eq.\ref{TPF:angle} is tilt that is obtained from angular dispersion and is the one most widely used in the context of terahertz generation \cite{hebling1996}. The second term in Eq.\ref{TPF:angle} describes pulse-front-tilt due to simultaneous spatial and temporal chirp. It indicates that if different colors are located at different transverse locations and each color arrives at a different time, then one obtains a pulse-front-tilt. This term has been described by prior work \cite{akturk2004}, but has not been examined in the context of terahertz generation. 

The third term in Eq.\ref{TPF:angle} is pulse-front-tilt introduced due to a finite radius of curvature and spatial chirp. While, prior work has suggested the relevance of the radius of curvature to pulse-front-tilt \cite{akturk2005}, here we provide an explicit expression. In the results and discussion section and Fig.\ref{fig4}, we specifically highlight the relevance of this term in conventional tilted-pulse-front setups.

\subsection{Polarization in the transverse momentum domain}

To solve Eq.\ref{e_thz_kx_a}, we first need to obtain an expression for $P_{THz}(\Omega, k_x,z)$, which is obtained by performing a Fourier transform of Eq.\ref{pol_thz_ST2}. This yields the following expression and associated effective parameters.

\begin{widetext}
\begin{subequations}
\begin{gather}
P_{THz}(\Omega,k_x,z) =
\frac{\varepsilon_0E_0^2\chi^{(2)} w}{2\tau}e^{-\frac{\Omega^2\tau^2}{8}}e^{-\frac{2\phi^2\Omega^2}{\tau^2+ w_0^2k_T^2\Omega^2}}e^{\frac{-(k_x-k_0\beta\Omega)^2[\frac{w^2}{2}-j\frac{\Omega w^2}{8cR(z)}]}{4}}e^{j(k_x-k_0\beta\Omega)x_0}\label{p_thz_kx}\\
\phi = \phi_0 + k_m z \label{phi_eff}\\
R^{-1}(z) = n(\omega_0)R_0^{-1}(z) +\frac{16c\zeta(z)k_T}{\tau^2w_0^2}\label{R_eff}\\
w^{-2} = w_0^{-2} + \frac{k_T^2\Omega^2}{\tau^2}\label{w_eff}\\
\beta = k_0\beta_0 -\frac{n(\omega_0)\omega_0\zeta(z)}{cR_0(z)} +\frac{2\zeta(z)\phi_0}{\frac{\tau_0^2w_0^2}{4} +\zeta(z)^2}\label{beta_eff}\\
x_0 = \frac{\phi}{k_T}\frac{k_T^2\Omega^2 w^2}{\tau^2}\label{x0}
\end{gather}
\end{subequations}
\end{widetext}

In Eq.\ref{p_thz_kx}, the first exponential term is identical to the first exponential term in Eq.\ref{pol_thz_ST2} and has already been discussed. The second term in Eq.\ref{p_thz_kx} is due to the group delay dispersion term $\phi_0 +k_m z$. From the third exponential term in Eq.\ref{p_thz_kx}, we see that the polarization is distributed about $k_x=k(\Omega)\text{sin}\gamma$. This indicates that the polarization term drives the radiation of terahertz waves at an angle $\gamma$ with respect to the pump direction, as expect. 

Various effective parameters in Eq.\ref{p_thz_kx} are delineated in Eqs.\ref{phi_eff} to \ref{x0}. Barring the net group delay dispersion term $\phi$ in Eq.\ref{phi_eff}, every term illustrates the importance of the role played by spatial-chirp and GVD-AD ($k_T$) terms in modifying the effective radius of curvature  $R$ (Eq.\ref{R_eff}), beam radius $w$ (Eq.\ref{w_eff}) and beam position $x_0$ (Eq.\ref{x0}). Furthermore, the modifications are more adverse for short durations $\tau$ or equivalently, large bandwidths. This point will be revisited repeatedly throughout the remainder of this paper.

\subsection{Terahertz spectra}

\subsubsection{Closed form expressions for $E_{THz}(\Omega,k_x,z)$}

While Eq.\ref{e_thz_kx_a} is valid in general , it requires numerical evaluation. However, it may be reduced to a closed-form expression in  frequency, transverse momentum  and longitudinal space, i.e.$(\Omega,k_x,z)$ as shown in Eq.\ref{e_thz_kx_c}(See Appendix for derivation).

\begin{widetext}
\begin{subequations}
\begin{gather}
E_{THz}(\Omega,k_x,z) = \frac{-j\Omega^2}{2\varepsilon_0 c^2 k_z}P_{THz}(\Omega,k_x,z)\bigg[\frac{e^{-j\Omega v_g^{-1}z}-e^{-\frac{\alpha(\Omega)z}{2\text{cos}\gamma}}e^{-jk_zz}}{\frac{\alpha(\Omega)}{2\text{cos}\gamma}+j\Delta k(\Omega,k_x)}\bigg] \label{e_thz_kx_c}\\
E_{THz}(\Omega,x,z) = \mathfrak{F}_x^{-1}\lbrace E_{THz}(\Omega,k_x,z)\rbrace\label{e_thz_x_0}\\
E_{THz}(t,x,z) = \frac{1}{2}\mathfrak{F}_t^{-1}\bigg[E_{THz}(\Omega,x,z)\Theta(\Omega) +E_{THz}(-\Omega,x,z)\Theta(-\Omega)\bigg]\label{e_thz_t_0}\\
\eta(z) =\frac{2\pi c\varepsilon_0}{F_{pump}\sqrt{\pi}w}\int_0^{\infty}\int_{-\infty}^{\infty}n_{THz}(\Omega)|E_{THz}(\Omega,x,z)|^2 dx d\Omega\label{eta_gen}
\end{gather} 
\end{subequations}
\end{widetext}

Upon obtaining Eq.\ref{e_thz_kx_c}, one may then obtain the spatial profile by taking an inverse spatial Fourier transform as shown in Eq.\ref{e_thz_x_0}. Subsequently, the real terahertz field $E_{THz}(t,x,z)$ may be obtained by an inverse temporal Fourier transform as delinated in Eq.\ref{e_thz_t_0}. Here $\Theta(\Omega)$ represents the Heaviside function. It delineates the fact that $E_{THz}(\Omega,x,z)$ calculated for $\Omega \geq 0$ has to be reflected about $\Omega=0$ in order to obtain the real field. By noting that $E_{THz}(-\Omega,x,z)=E_{THz}^{*}(\Omega,x,z)$, it is evident that $E_{THz}(t,x,z)$ shall be a real valued function. Equation \ref{eta_gen} corresponds to optical-to-terahertz conversion efficiency $\eta$ obtained by integrating the spectral intensity over space and angular frequency. By Parseval's theorem, this could also be calculated in transverse momentum $k_x$ and/or temporal $t$ domains (See Appendix).

The validity of Eq.\ref{e_thz_kx_c} maybe verified by noticing that, in the absence of loss, it reduces to the all familiar form proportional to the $\text{sinc}$ function presented in Eq.\ref{e_thz_kx_noloss}.

\begin{gather}
E_{THz}(\Omega,k_x,z) = \frac{-j\Omega^2}{2\varepsilon_0 c^2 k_z}P_{THz}(\Omega,k_x,z)\text{sinc}\left(\frac{\Delta k z}{2}\right)\label{e_thz_kx_noloss}
\end{gather}

Equation \ref{e_thz_kx_c} accounts for effects of loss, dispersion in both optical and terahertz frequency ranges, spatio-temporal distortions such as GVD-AD and spatial chirp as well as spatial walk-off between terahertz and optical pulses. Importantly, since no assumption is made on limiting the range of $k_z$, this expression \textit{does not contain paraxial approximations} and is thus generally valid for all pump beam sizes and propagation distances (i.e near and far-field). Furthermore, Eq.\ref{e_thz_kx_c} is equally valid for conditions of low and high absorption. It can be thus used quite generally to obtain meaningful predictions of terahertz efficiency, field strength, frequency and spatio-temporal profiles in the undepleted limit.

\subsubsection{Closed form expressions for $E_{THz}(\Omega,x,z)$}

In order to obtain physical intuition for the spatio-temporal properties of the generated terahertz radiation,closed-form expressions of the terahertz spectrum $E_{THz}(\Omega,x,z)$ is desirable. 

However, due to the presence of functions of $k_x$ in the denominator of Eq.\ref{e_thz_kx_c}, a general closed-form expression for $E_{THz}(\omega,x,z)$ appears intractable. However, one may be obtained if the following conditions are satisfied.

If the transverse-momentum distribution $P_{THz}(\Omega,k_x,z)$ in Eq.\ref{p_thz_kx} is highly localized about $k_x=k(\Omega)\text{sin}\gamma$ or when the phase-mismatch $\Delta k =0$, then the $k_x$ dependencies in the denominator of Eq.\ref{e_thz_kx_c} maybe eliminated. Formally, this constraint maybe delineated by setting $\Delta k \ll \alpha$ at $k_x= k(\Omega)\text{sin}\gamma \pm 2\sqrt{2}/w$, or when the transverse momentum distribution in Eq.\ref{p_thz_kx} reduces to $e^{-1}$ of its value. Furthermore, accounting for the $\Omega$ dependence of $w$ as given by Eq.\ref{w_eff} , we obtain the following constraints :

\begin{subequations}
\begin{gather}
w_0 > \frac{4\sqrt{2}\text{sin}\gamma}{\alpha}\label{constraint_a}\\
4w_0^2k_T^2\tau^{-4}\ll 1\label{constraint_2}
\end{gather}
\end{subequations}
 
Equations \ref{constraint_a} and \ref{constraint_2} suggest that while on one hand beam sizes need to be large enough , they must also satisfy the condition that $2k_Tw_0/\tau^2$ be small enough. In ensuing sections we will establish that very important experimental situations fulfill these conditions, thus making the forthcoming expressions relevant to practical situations.

Since Eq.\ref{constraint_a} requires transverse momentum distributions to be highly localized (possess large beam sizes), $k_z(\Omega)$  maybe approximated \textit{paraxially} as $k_z=k_{THz}\text{cos}\gamma -\frac{k_x'^2}{2k_{THz}\text{cos}\gamma} -k_x'\text{tan}\gamma$, where $k_x'$ is a slight transverse momentum variation about $k_x = k(\Omega)\text{sin}\gamma$. Invoking the above , one obtains the following expression for $E_{THz}(\Omega,x,z)$ in Eq.\ref{e_thz_x_1} upon taking the inverse Fourier transform of Eq.\ref{e_thz_kx_c} in the paraxial limit.

\begin{widetext}
\begin{subequations}
\begin{gather}
E_{THz}(\Omega,x,z) = \frac{-j\Omega}{n_{THz}(\Omega)\alpha(\Omega) c}\frac{\chi^{(2)}E_0^2\sqrt{2\pi}}{\tau}e^{-j\Omega v_g^{-1} z}e^{-jk(\Omega)\text{sin}\gamma x}
\bigg[e^{-\frac{\Omega^2\tau_1^2}{8}}e^{-x^2\bigg(2w_0^{-2} +\frac{j\Omega}{2cR(z)}\bigg)}\nonumber\\
-e^{-\frac{\alpha(\Omega)z}e^{-\frac{\Omega^2\tau_2^2}{8}}{2\text{cos}\gamma}}e^{-(x-z\text{tan}\gamma)^2\bigg(2w_0^{-2} +\frac{j\Omega}{2cR(z)}\bigg)}\bigg] \label{e_thz_x_1}\\
\tau_1 =\tau\left[1+\frac{16x^2k_T^2}{\tau^4} +\frac{16\phi^2}{\tau^4}\right]^{1/2}\label{tau_1}\\
\tau_2 =\tau\left[1+\frac{16(x-z\text{tan}\gamma)^2k_T^2}{\tau^4} +\frac{16\phi^2}{\tau^4}+\frac{4\beta_T}{\tau^2}\right ]^{1/2}\label{tau_2}
\end{gather}
\end{subequations}
\end{widetext}

Examining Eq.\ref{e_thz_x_1}, the complex exponential factors outside the square brackets clearly delineate a terahertz pulse propagating at an angle $\gamma$ relative to the pump pulse. Further, note that the spectral intensity is inversely proportional to the absorption coefficient. The first term inside the square brackets represents the source term driven by the nonlinear polarization term in Eq.\ref{pol_thz_ST2}. The effective duration of this nonlinear polarization term is given by $\tau_1$ as defined in Eq.\ref{tau_1}. As is evident, $\tau_1$ has a transverse variation, which delineates spatial chirp of the generated terahertz pulse proportional to the GVD-AD term $k_T$. The extent of this transverse variation is accentuated at larger bandwidths or shorter $\tau$. Furthermore, the effective terahertz pulse duration is enlarged by the effective GDD $\phi$ (Eq.\ref{phi_eff}). The $x_0$ term defined in Eq.\ref{x0} does not feature in the above equations due to the assumption of small $|k_T|w_0/\tau^2$, which causes $x_0\approx 0$ according to Eq.\ref{x0}. From Eq.\ref{e_thz_x_1}, a finite radius of curvature $R(z)$ is also imparted to the terahertz spectrum which is non-zero even for an infinite pump radius of curvature due to spatio-temporal coupling effects as delineated in Eq.\ref{R_eff}. 

The second term within square-brackets in Eq.\ref{e_thz_x_1} on the subsequent line represents the propagating terahertz wave. The total field is thus a difference of the nonlinear polarization term and the propagating terahertz wave. Here, notice that since the terahertz pulse propagates at an angle $\gamma$ with respect to the pump beam, the spatial profile is distributed about the line $x-z\text{tan}\gamma=0$. The difference in the spatial distributions is what creates the walk-off effect. The effective duration of the propagating component in Eq.\ref{e_thz_x_1} is given by $\tau_2$ as defined in Eq.\ref{tau_2}. Naturally, the distribution of $\tau_2$ spatially is different from $\tau_1$ but in addition, it increases due to group velocity dispersion in the terahertz region $\beta_T$. 

\subsubsection{Spatial terahertz profiles}

\begin{figure}
\centering
\includegraphics[scale=0.45]{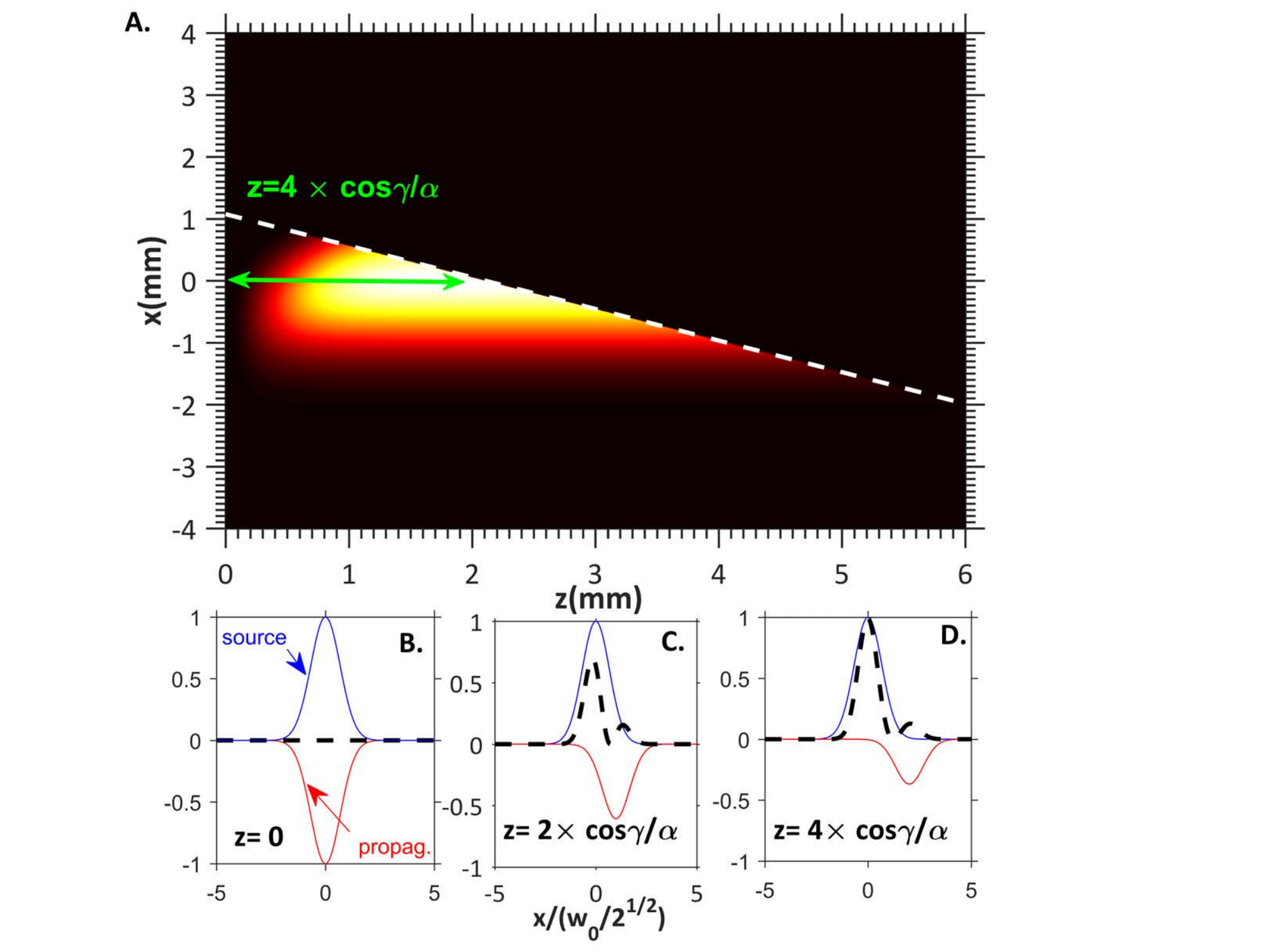}
\caption{Understanding the spatial distribution of terahertz energy in tilted-pulse-front setups. (a)Distribution of terahertz fluence in the $x-z$ plane of the crystal. The boundary of the crystal is represented by the white-dotted line. The asymmetry in the energy distribution is due to an interplay between the geometry and the effects of absorption and walk-off.(b) The distribution of $E_{THz}(\Omega)$ in $x$ is the difference of two profiles centered at $x=0$ and $x=z\text{tan}\gamma$ evident from Eq.\ref{e_t_x_1}. At z=0, the two terms cancel out. (c). For larger $z=2\text{cos}\gamma/\alpha$, the displaced and attenuated second term produces an asymmetric total transverse profile of larger strength. This is the case for $z<1$mm in (a).(d) For large enough $z$, the second term in Eq.\ref{e_thz_x_1} is attenuated sufficiently and displaced faraway from $x=0$ to result in the $E_{THz}(\Omega,x,z$ reaching a maximum value. This is the case for $z>2$mm in (a).}
\label{fluence_map}
\end{figure}

Using Eqs.\ref{e_thz_x_1}-\ref{tau_2}, we proceed to understand the spatial distribution of the generated terahertz radiation in this section. In Fig.\ref{fluence_map}(a), we plot the terahertz fluence (or spectral intensity integrated over all terahertz angular frequencies) for a lithium niobate crystal with $w_0=2$mm and $\tau_0=500$fs at T=300 K in the $x-z$ plane. The parameters from Table.\ref{param_list} are used for calculations.

The white-dotted line in Fig.\ref{fluence_map}(a) delineates the crystal boundary and the optical pump beam is centered about $x=0$. The fluence distribution depicts an asymmetric beam profile, which may be understood by examining Eq.\ref{e_thz_x_1} and Figs.\ref{fluence_map}(b)-(d).As previously discussed, in Eq.\ref{e_thz_x_1}, the terahertz spectrum $E_{THz}(\Omega,x,z)$ is the difference between two transverse spatial profiles. The first term inside the square brackets of Eq.\ref{e_thz_x_1} is centered about $x=0$ and maybe viewed as \textit{source} (blue curve in Fig.\ref{fluence_map}(b)-(d)) or near-field term(\citep{bakunov2008}. The second term inside the square brackets of Eq.\ref{e_thz_x_1}, is centered about $x=z\text{tan}\gamma$ and corresponds to the radiated or \textit{propagating term} (red curve in Fig.\ref{fluence_map}(b)-(d)), further evident by the fact that it suffers attenuation as it propagates. The total transverse profile is the difference of these two profiles.

As illustrated in Fig.\ref{fluence_map}(b), at $z=0$, the two transverse profiles are identical and cancel each other out, resulting in the expected value of $E_{THz}(\Omega,x,z=0)=0$ (black-dotted curve in the insets of Fig.\ref{fluence_map}). As $z>0$, the propagating term is displaced to positive $x>0$, which results in imperfect cancellation to produce an asymmetric total profile as seen by the black-dotted curve in  Fig.\ref{fluence_map} (b) and also in the fluence map for $z<1$mm. A growth in intensity from $0$ is seen due to the attenuation of the propagation term in Eq.\ref{e_thz_x_1}, which then reaches a maximum beyond the absorption length $\sim 2\text{cos}\gamma/\alpha$.  While for small $z$, the asymmetry in the transverse profile is a result of the physics of the nonlinear process described in Figs.\ref{fluence_map}(b)-(d), for large $z$, only parts of the beam for $x<0$ lie within the crystal.

\subsubsection{Efficiency variation with length}
Insights into the optimal interaction length can be gleaned by evaluating the conversion efficiency $\eta$ from Eq.\ref{eta_gen} with the terahertz spectral profile $E_{THz}(\Omega,x,z)$ obtained from Eq.\ref{e_thz_x_1} as follows :

\begin{subequations}
\begin{gather}
\eta(z) = \frac{F_{pump}{\chi^{(2)}}^2}{2\pi\varepsilon_0n_{THz}n_{IR}^2c^3}F(\Omega)\nonumber\\
\times \bigg[\left(1-e^{-\frac{\alpha z}{2\text{cos}\gamma}}\right)^2 + 2e^{-\frac{\alpha z}{2\text{cos}\gamma}}\left(1-e^{-\frac{z^2\text{tan}^2\gamma}{w_0^2}}\right)\bigg]\label{eta}\\
F(\Omega)=\bigg[\int_0^{\infty}\frac{\Omega^2e^{-\frac{\Omega^2\tau^2}{4}}}{\alpha(\Omega)^2\sqrt{1+\frac{k_T^2\Omega^2w_0^2}{\tau^2}}}d\Omega\bigg]\label{kernel}
\end{gather}
\end{subequations}

The first $z$ dependent term inside the square brackets of Eq.\ref{eta}, represents saturation of conversion efficiency due to absorption. The second term represents limitations due to walk-off. For beam sizes $w_0\ll \alpha^{-1}$, the second term dominates and the system would be walk-off limited. The optimal interaction length would thus be $z_{opt}\sim w_0/\text{tan}\gamma$. For large beam sizes $w_0\gg\alpha^{-1}$, the first term would dominate and the system would be primarily absorption limited with $z_{opt}\sim 2\text{cos}\gamma/\alpha$. In practice, cascading effects reduce the interaction length further, which are not considered in this analytic study. For lithium niobate, at room temperature, conversion efficiencies are lower and the situation is closer to the undepleted case.

While Eq.\ref{kernel} does not have an explicit form in general, it may be evaluated for constant $\alpha$ and $|k_T|w_0/\tau^2 \ll 1$. For large enough $z$, the conversion efficiency $\eta$ maybe approximated by the following :

\begin{subequations}
\begin{gather}
\eta \approx \frac{F_{pump}{\chi^{(2)}}^2}{\sqrt{\pi}\varepsilon_0n_{THz}n_{IR}^2c^3}\tau^{-3}\bigg[1-\frac{3k_T^2w_0^2}{\tau^4}\bigg]\label{eta_approx}\\
\tau_{opt}\approx 1.62(|k_T|w_0)^{1/2}\label{tau_opt}
\end{gather}
\end{subequations}

Note, the reduction of conversion efficiency for large $k_T$(GVD-AD) and beam radii. In the absence of GVD-AD, the conversion efficiency should improve with pump bandwidth or shorter $\tau$. However, due to the effects of GVD-AD, an optimal pulse duration $\tau_{opt}\approx 1.62(k_Tw_0)^{1/2}$ exists. 

\subsubsection {Terahertz frequency}

Maximizing Eq.\ref{kernel} with respect to terahertz angular frequency $\Omega$, one may obtain the following expression for the central terahertz frequency $\Omega_{max}$ for the case when $k_T2w_0/\tau^2 \ll 1$ as shown below :

\begin{gather}
\Omega_{max} \approx \frac{2}{\tau}\bigg[1-2\frac{k_T^2w_0^2}{\tau^4}\bigg]\label{freq_1}
\end{gather}

Once again, the reduction of average frequency for larger bandwidths and beam radii is evident for systems with non-zero GVD-AD. This trend is consistent with experimental observations for tilted-pulse-front experiments in lithium niobate.

\subsubsection{Validity of analytic expressions of $E_{THz}(\Omega,x,z)$}

\begin{figure}
\centering
\includegraphics[scale=0.325]{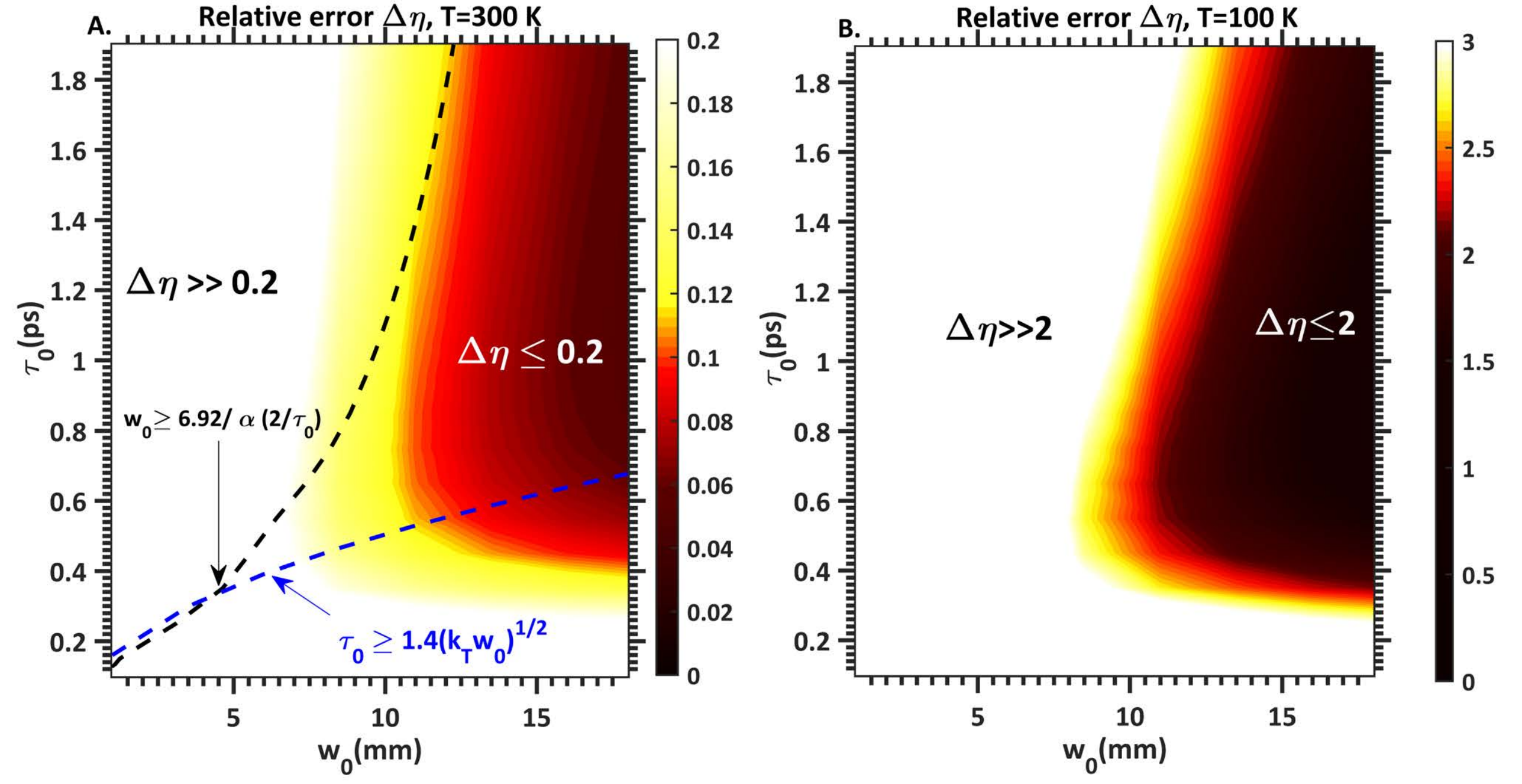}
\caption{Comparison of conversion efficiencies $\eta$ calculated from expressions for spectra in Eq.\ref{e_thz_x_1} and Eq.\ref{e_thz_kx_c} for a lithium niobate crystal. (a) At T=300 K, $\Delta\eta\leq 20\%$ if constraints in Eq.\ref{constraint_a} (Black-dotted) and Eq.\ref{constraint_2} (Blue-dotted) are satisfied. (b) At T=100 K, relative error is large predominantly due to violation of constraint in Eq.\ref{constraint_a} due to reduced absorption at cryogenic temperatures.}
\label{rel_err}
\end{figure}

While Eq.\ref{e_thz_kx_c} is generally valid and does not incur significantly more computational cost compared to evaluating Eq.\ref{e_thz_x_1}, the latter is appealing due to being explicit and more intuitive. We have already laid out the conditions when Eq.\ref{e_thz_x_1} can be employed in Eqs.\ref{constraint_a}-\ref{constraint_2}. Here, we plot the relative error in conversion efficiency $\Delta\eta$ between Eqs.\ref{e_thz_x_1} and \ref{e_thz_kx_c} for terahertz generation in lithium niobate in Fig.\ref{rel_err}. Parameters for the calculations are provided in Table.2. 

From Fig.\ref{rel_err}(a), it can be seen that for T=300 K or in the large absorption limit, the closed-form expression in Eq.\ref{e_thz_x_1} is within $20\%$ of the general expression in Eq.\ref{e_thz_kx_c} for regions approximately bounded by the constraints in Eq.\ref{constraint_a} and \ref{constraint_2}. They are particularly accurate for $\approx cm$ beam sizes, making them attractive for analyzing high energy terahertz generation setups. For instance, at $\tau_0\approx 0.5$ps ,$f_{THz}\approx 0.5$ THz, we obtain the threshold beam radius $w_0\approx 7$ mm from Eq.\ref{constraint_a}, which is consistent with Fig.\ref{rel_err}(a). However, Eq.\ref{e_thz_x_1} is less efficacious for very small beam radii or short pulse durations.
 
We set $\alpha=\alpha(2/\tau_0)$ in Eq.\ref{constraint_a}, since the central terahertz frequency is approximately $\Omega_{max}\approx 2/\tau_0$ (Eq.\ref{freq_1}). Therefore, the threshold beam size is larger for longer pump pulses due to the reduced terahertz frequency and hence smaller absorption.   

Naturally, the discrepancy between Eqs.\ref{e_thz_x_1} and \ref{e_thz_kx_c} is much larger for T=100 K due to a much smaller absorption coefficient of lithium niobate at cryogenice temperatures. For 0.5 THz, the absorption coefficient at T=100 K $\approx 130/$ cm which is $>7$ times smaller compared to that at T=300K ($\approx 960/$cm). This translates to the critical beam size for low relative error at T=100 K being about 5.5 cm, which is out of bounds in the parameter space depicted in Fig.\ref{rel_err}(b). 

\subsection{Temporal profiles}

Here, we present closed-form expressions for terahertz transients when the constraints supplied by Eqs.\ref{constraint_a}-\ref{constraint_2} are satisfied. This may be obtained by taking an inverse Fourier transform in the temporal domain of the expression provided by Eq.\ref{e_thz_x_1}. To obtain, a closed-form expression, the dispersive properties of the absorption coefficient are neglected and $\alpha$ is set to $\alpha(2/\tau_0)$, since the central terahertz frequency is roughly given by $2/\tau_0$ (Eq.\ref{freq_1}).The results are presented in Eqs.\ref{e_t_x_1}-\ref{t_2} below.

\begin{widetext}
\begin{subequations}
\begin{gather}
E_{THz}(t,x,z) = \frac{-8\pi\chi^{(2)}E_0^2}{n_{THz}\alpha(2/\tau_0) c\tau}
\bigg[\frac{1}{\tau_1^3}e^{-2x^2w_0^{-2}}t'e^{-\frac{2t'^2}{\tau_1^2}}
-\frac{1}{\tau_2^3}e^{-\frac{\alpha_0 z}{2\text{cos}\gamma}}e^{-2(x-z\text{tan}\gamma)^2w_0^{-2}}t"e^{-\frac{2t"^2}{\tau_2^2}}\bigg]\label{e_t_x_1}\\
t' = t-x\text{sin}\gamma v_{THz}^{-1} -z\text{cos}\gamma v_{THz}^{-1} + \frac{x^2}{2cR(z)}\label{t_1}\\
t" = t-x\text{sin}\gamma v_{THz}^{-1} -z\text{cos}\gamma v_{THz}^{-1} + \frac{(x-z\text{tan}\gamma)^2}{2cR(z)}\label{t_2}
\end{gather}
\end{subequations}
\end{widetext}

Firstly, note that the intensity along the plane defined by $t'=t-x\text{sin}\gamma v_{THz}^{-1}-z\text{cos}\gamma v_{THz}^{-1}$ is constant due to an obliquely propagating terahertz pulse. Secondly, the finite radius of curvature slightly modifies this plane to a curved surface as is evident in Eq.\ref{t_1}. Secondly, as already delineated in Eqs.\ref{tau_1}-\ref{tau_2}, the pulse duration of the terahertz transients varies along transverse spatial dimension $x$. The degree of spatial variation increases with pump bandwidth or shorter $\tau$. 

\begin{figure*}
\centering
\includegraphics[scale=0.425]{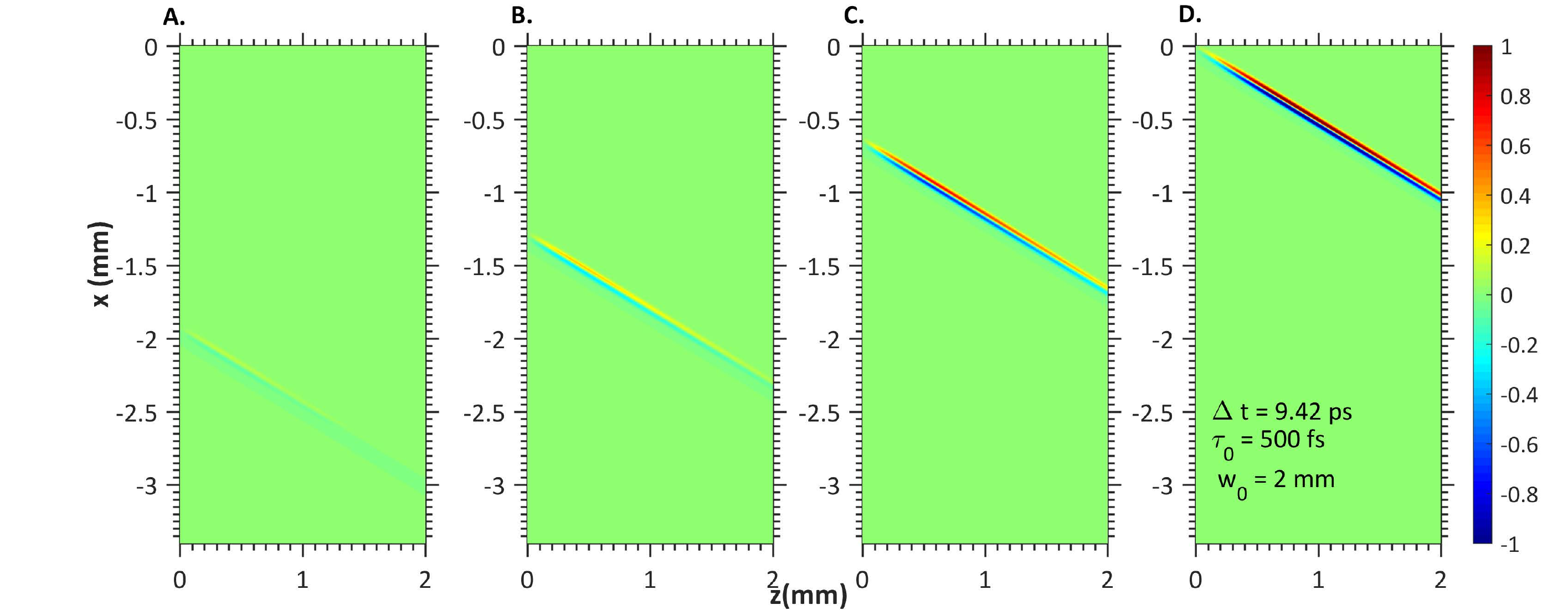}
\caption{Spatio-temporal evolution of terahertz transients for a $\tau_0=500$ fs, $w_0=2$mm beam radius in lithium niobate at T=300 K. The propagation of the terahertz transient at an angle $\gamma \approx 63^\circ$ is evident in panels (a)-(d). A continuous growth in intensity is seen. Due to the large duration of the pump pulse, spatial inhomogeneities are relatively low.}
\label{e_t_500}
\end{figure*}

\begin{figure*}
\centering
\includegraphics[scale=0.4]{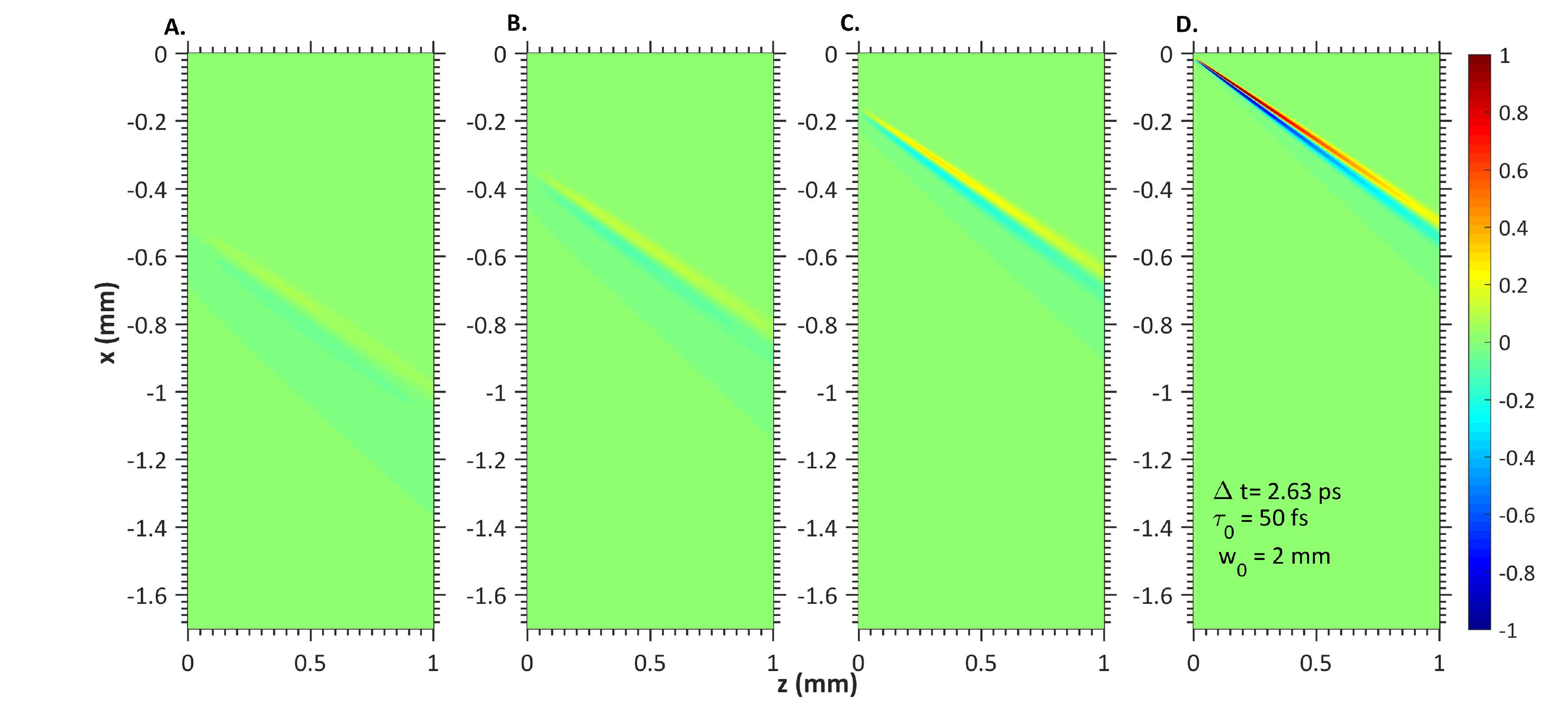}
\caption{Spatio-temporal evolution of terahertz transients for a $\tau=50$fs, $w_0=2$mm beam radius in lithium niobate at T=300 K. Due to the short duration of the pump pulse, spatial inhomogeneities are obviously present in (d).}
\label{e_t_50}
\end{figure*}

We employ Eq.\ref{e_t_x_1} to plot the evolution of the terahertz electric field in lithium niobate at T = 300K for for two different durations $\tau=500$fs in Fig.\ref{e_t_500} and $\tau=50$fs in Fig.\ref{e_t_50}. As can be seen in Fig.\ref{e_t_500}, the field grows as it propagates at an angle $\gamma = 63^\circ$ with respect to the pump, while evolving into a single-cycle terahertz field. Across the tilt-plane, the duration of the pulse does not vary noticeably, with relatively uniform properties.

However, for $\tau_0=50$fs, in Fig.\ref{e_t_50}, the field is seen to grow as was the case for Fig.\ref{e_t_500} but the degree of asymmetry in pulse duration along the tilt-plane is greater due to the effects of GVD-AD as expected from Eqs.\ref{tau_1} and \ref{tau_2}.

By inspecting Eq.\ref{e_t_x_1}, it is clear the peak field $E_{THz,max}$ occurs at the maxima of the function $t'e^{-2t'^2/\tau^2}$ which is at $t'=\tau/2$. Thus the peak field is given by the following expression in Eq.\ref{e_max} :

\begin{gather}
|E_{THz,max}(\tau_0)| \leq \frac{4\pi\chi^{(2)}E_0^2e^{-\frac{1}{2}}}{n_{THz}\alpha(2/\tau_0) c\tau_0^3} \label{e_max}
\end{gather}

\subsubsection{Validity of closed-form expression for transients}

\begin{figure}
\centering
\includegraphics[scale=0.4]{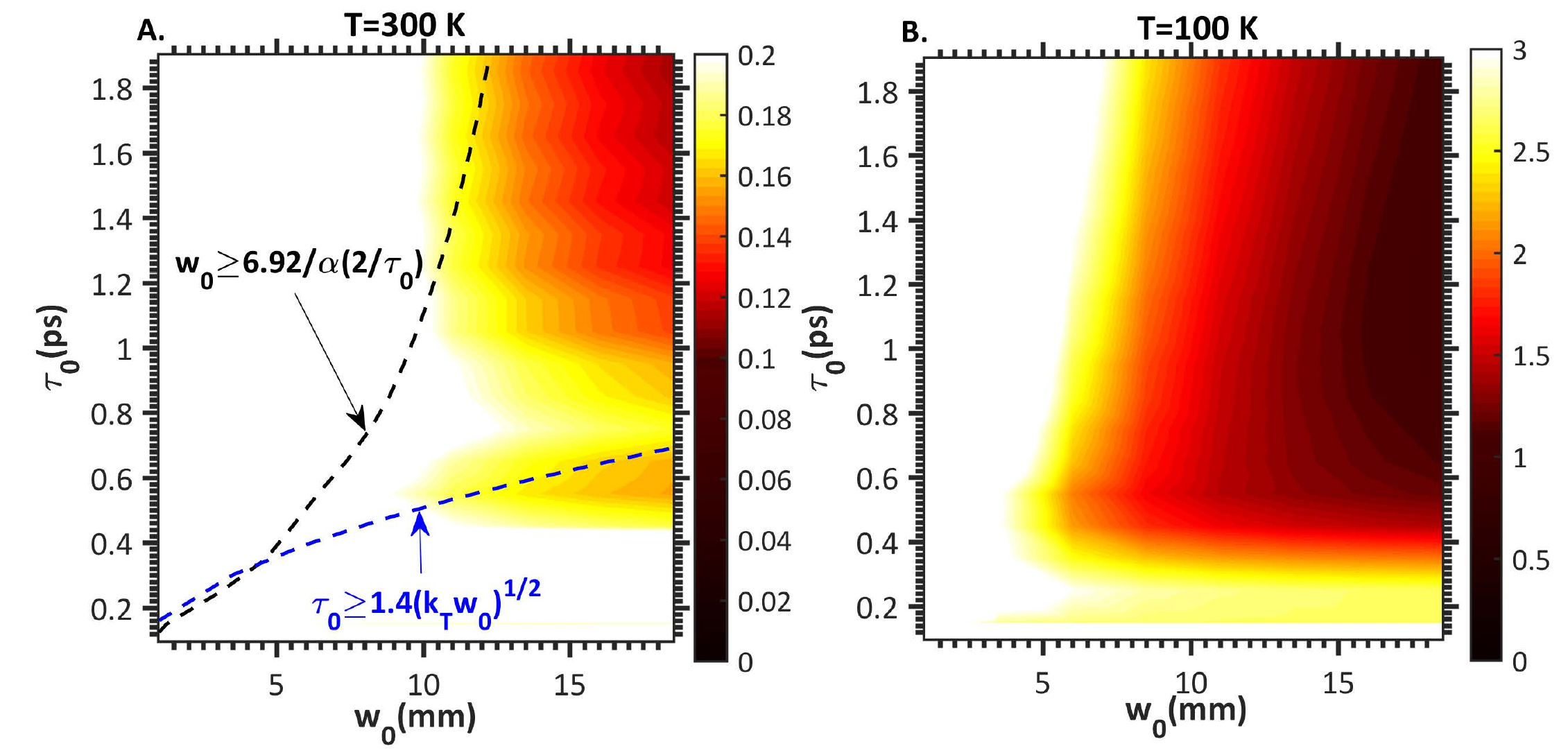}
\caption{Relative error of peak field calculations between Eqs.\ref{e_t_x_1} and \ref{e_thz_t_0}. Consistent with Fig.\ref{rel_err}, the relative error is $\leq 20\%$ at T=300 K upon satisfying the constraints in Eqs.\ref{constraint_a} (black-dotted) and \ref{constraint_2} (blue-dotted). (b) At T=100 K, the relative error is large, analogous to Fig.\ref{rel_err}(b) due to violation of Eq.\ref{constraint_a} resulting from small absorption at cryogenic temperatures.}
\label{rel_err_emax}
\end{figure}

In Fig.\ref{rel_err_emax}, we test the proximity of the bounds provided by the peak field in Eq.\ref{e_max} to those obtained from the general expressions in Eqs.\ref{e_thz_kx_c}-\ref{e_thz_t_0} for lithium niobate . Similar to Fig.\ref{rel_err}, the agreement is within $20\%$ at T=300 K for regions bounded approximately by the constraints in Eqs.\ref{constraint_a} -\ref{constraint_2}. However, just as in the case of Fig.\ref{rel_err}(b), the relative error is large for T=100 K in the parameter space of $w_0-\tau_0$ depicted. Hence, the closed form expression of transients are also more useful at room temperature for high energy tilted-pulse-front setups (i.e large beam radii). In Table.\ref{valid_list}, we summarize the various equations and their domains of applicability.

\begin{table*}[ht]
\begin{tabular}{l c r}
\hline
Equation & Description & Validity \\
\hline
Eqs.\ref{e_thz_kx_c}-\ref{e_thz_t_0} & Closed form for $E_{THz}(\omega,k_x,z)$ &  High(Quantitative) and low absorption (Qualitative).  \\

Eqs.\ref{e_thz_x_1},\ref{eta} & Closed-form for $E_{THz}(\Omega,x,z)$  & High (Quantitative) absorption for Eqs.\ref{constraint_a}-\ref{constraint_2}\\

Eqs.\ref{e_t_x_1}, \ref{e_max} & Closed-form for $E_{THz}(t,x,z)$  & High (Quantitative) absorption for Eqs.\ref{constraint_a}-\ref{constraint_2}.\\
\hline
\end{tabular}
\caption{\label{valid_list} Summary of equations and their domains of validity}
\end{table*}

\section{Results and discussion}
\begin{table*}[ht]
\begin{tabular}{l c r}
\hline
Parameter & Symbol & Value \\
\hline
Second order nonlinear coefficient & $\chi^{(2)}$ & 336 pm/V  \\

Central pump wavelength &$\lambda_0$   & 1.03 $\mu$m   \\

Optical phase index & $n_{IR}$ & 2.17  \\

Optical group index  & $n_g$ & 2.25 \\

Terahertz phase index (Eqs.\ref{eta},\ref{eta_approx},\ref{e_t_x_1})  & $n_{THz}$& 4.75 (T = 100 K)\\
&& 4.95 (T = 300 K)\\

GVD-AD & $k_T$ & $-1.3\times 10^{-23}$ s$^2$/m \\

Full Terahertz index dispersion (Eqs.\ref{e_thz_kx_c},\ref{e_thz_x_1}) & $n_{THZ}(\Omega)$ & \cite{palfalvi2005}\\
Full Terahertz absorption dispersion (Eqs.\ref{e_thz_kx_c},\ref{e_thz_x_1}) & $\alpha(\Omega)$&\cite{fulop2011}\\
Peak Intensity  & $I_{max}$ & 50 GW/cm$^2$\\
Fluence & $F_{pump}$ & $I_{max}\sqrt{\pi/2}\tau$ J/m$^2$\\     
\hline
\end{tabular}
\caption{\label{param_list} List of parameters used in calculations.}
\end{table*}
In this section, we analyze important features of terahertz generation with tilted-pulse-fronts in lithium niobate. We first illustrate the importance of spatial-chirp and radius of curvature from Eq.\ref{TPF:angle} to the tilt angle $\gamma$. Subsequently, we  employ Eqs.\ref{e_thz_kx_c}-\ref{e_thz_t_0} to  evaluate terahertz spectra $E_{THz}(\Omega,x,z)$, conversion efficiency $\eta$ as well as peak electric fields at T=300 K and T= 100 K for various values of beam radius $w_0$ and pump pulse duration $\tau_0$. We compare cases with and without GVD-AD $k_T$ to illustrate its effects on terahertz frequency, conversion efficiency and peak electric field. We find that for larger beam radii and shorter pump pulse durations, there is a decrease in all these quantities due to the detrimental effects of GVD-AD. Overall the undepleted models provide good qualitative predictions and understanding of terahertz generation using tilted-pulse-fronts. However, while quantitative predictions for T=300 K are reasonable , those at T=100 K are overestimated. Although absorption reduces significantly at cryogenic temperatures, cascading effects induce limitations which necessitates full depleted calculations for these conditions.

\subsection{Tilt angle due to spatial-chirp and radius of curvature}

\begin{figure}
\centering
\includegraphics[scale=0.45]{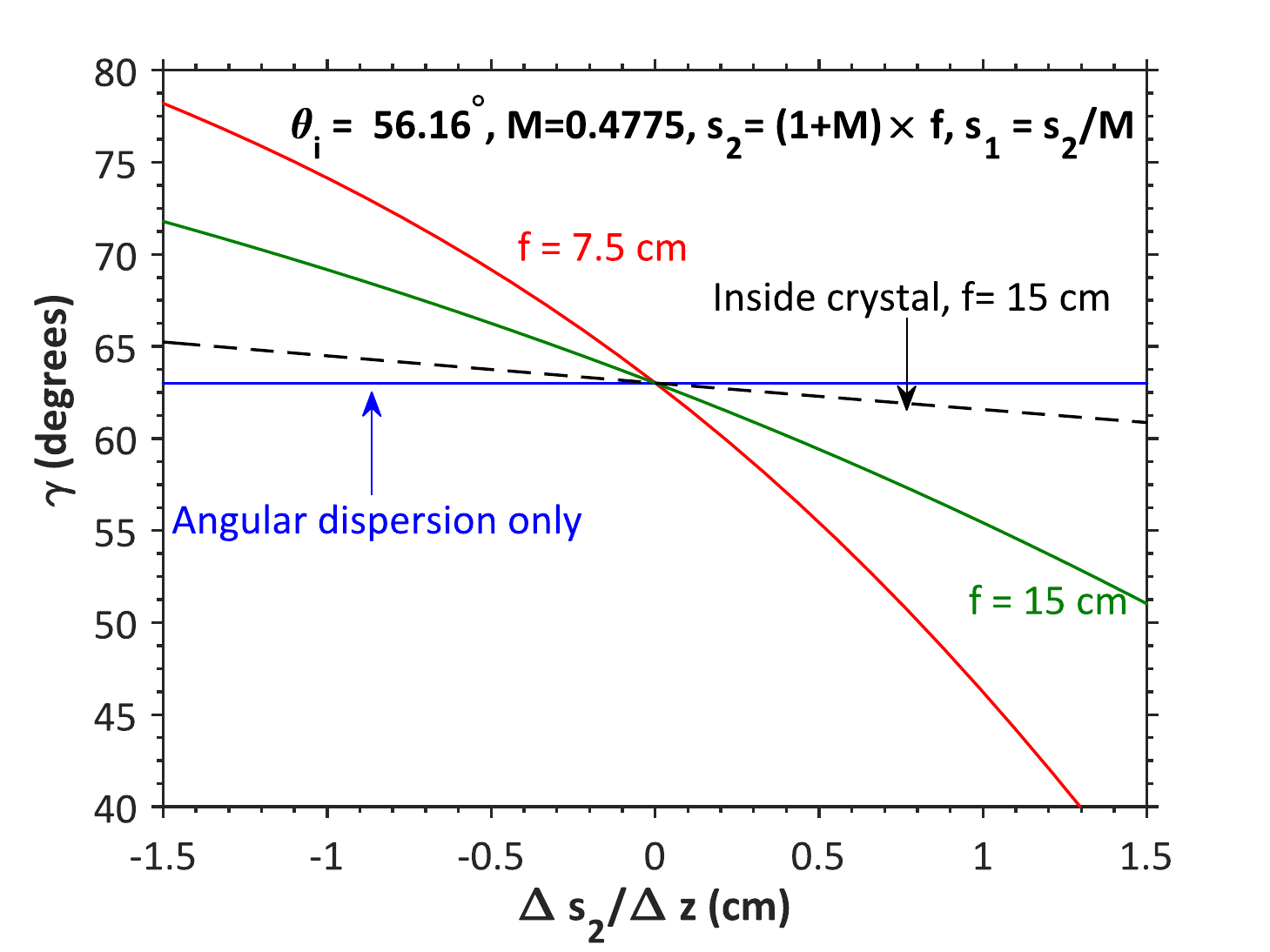}
\caption{Variation of pulse-front-tilt angle $\gamma$ just inside a lithium niobate crystal with lens-to-crystal distance variations $\Delta s_2$. Angular dispersion(blue) does not change with $s_2$ as expected. Variations of pulse-front-tilt due to spatial-chirp $\zeta(z)$ and radius of curvature $R(z)$ (third term, Eq.\ref{TPF:angle}) however produces significant variations in tilt-angle. The variations are more severe for short focal length of $f=7.5$cm(red) compared to longer focal length of $f=15$cm(green). Variation of tilt-angle due to spatial-chirp and radius of curvature inside the crystal is negligible due to reduction in growth of $\Delta\zeta(z)$ due to reduced angular dispersion (reduces by a factor $n{\omega_0}$ due to Snell's law) as well as slower change of radius of curvature.}
\label{fig4}
\end{figure}

The importance of the contribution of the radius of curvature and spatial-chirp to the tilt-angle is particularly evident in considering the effect of the lens-to-crystal distance $s_2$ (see Fig.\ref{fig1}) on conversion efficiency. Experimentally, the displacement of the crystal from the optimal imaging distance $s_2$ results in a dramatic loss in conversion efficiency \cite{ravi15}. In Fig.\ref{fig4}, we plot the tilt-angle of the pump beam \textit{just inside} the crystal for various values of $s_2$ using Eq.\ref{TPF:angle}. A tilted-pulse-front setup with grating of $p=1500$ lines/mm, grating incidence angle $\theta_i =56.16^\circ$, lens focal length $f$, magnification $M =0.4775$, $s_2 = (1+M)f$ and  $s_1 = s_2/M$  was assumed. Further, the pulse properties corresponded to a duration $\tau_0=50$ fs and beam radius $w_0=1$ mm. The values of $\zeta,\beta_0,R_0(z)$ were calculated using dispersive ray-pulse matrices, following Martinez \cite{martinez1988}.

As expected,the angular dispersion of the beam does not change after the focus (which lies before the crystal) as is evident by the flat blue line in Fig.\ref{fig4}. Secondly, the contribution to the tilt-angle from the spatio-temporal chirp is found to be infinitesimally small (on the order of $10^{-6}$ even for group delay dispersion $\phi_0$ on the order of $5\times10^5\text{fs}^2$ (for $\tau_0=50$fs). However, the contribution from the final term of Eq.\ref{TPF:angle} comprising of spatial-chirp and the radius of curvature is responsible for a significant change in tilt-angle as can be seen in Fig.\ref{fig4}. This variation arises from changes in both spatial chirp $\Delta\zeta = n(\omega_0)\beta_0 \Delta s_2$ as well as changes in the radius of curvature with displacement of the crystal $\Delta s_2$. As Fig.\ref{fig4} suggests, the use of longer focal lengths (green curve) reduces the sensitivity of the setup to displacements $\Delta s_2$, making it easier  to obtain optimal performance.

Inside the crystal, this term has a much smaller impact since the variation of  spatial chirp with distance $z$ is smaller due to reduced angular dispersion inside the material (by Snell's law), i.e. $\Delta\zeta =\beta_0z$. Furthermore, the variation of the radius of curvature is also reduced. Thus, the impact on interaction length within the crystal due to varying tilt-angle is much lesser (black,dashed line in \ref{fig4}).If a near collimated beam (with $R=\infty$ can be produced, then the impact of this effect will of course be negligible. Thus, the use of telescopic setups for imaging may indeed prove to be advantageous as prior experiments \cite{hirori2011} seem to suggest. 

\subsection{Spectra and frequency}

\begin{figure}
\centering
\includegraphics[scale=0.425]{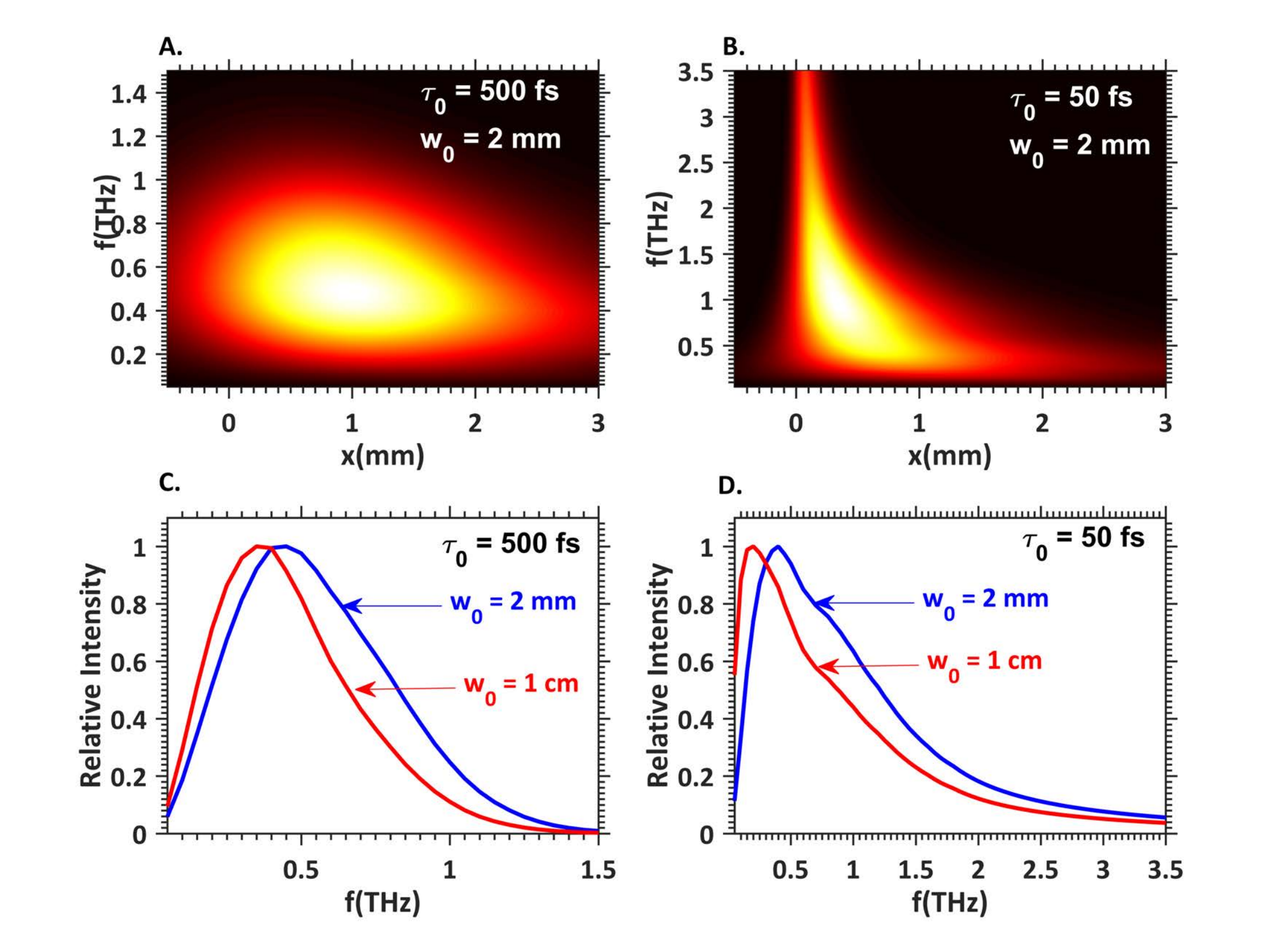}
\caption{Terahertz spectra obtained from Eqs.\ref{e_thz_kx_c}-\ref{e_thz_x_0} in lithium niobate at T=300 K. (a) Spatial distribution of the terahertz spectrum for a pump pulse duration of $\tau_0=500$fs and $w_0=2$ mm shows relatively homogeneous properties consistent with transients in Fig.\ref{e_t_500}.(b) Spatial distribution of the terahertz spectrum for a pump pulse duration of $\tau_0=50$ fs and $w_0=2$ mm shows significant asymmetry analogous to Fig.\ref{e_t_50}.(c) Average spectra $\int|E_{THz}(\Omega,x,z)|^2dx$ for varying beam sizes shows the reduction in frequency bandwidth for larger beam radii due to the effects of group-velocity dispersion due to angular dispersion (GVD-AD).(d) Average spectra for $\tau_0=50$fs shows even greater reduction of frequency due to enhancement of GVD-AD effects at short durations.(See Eq.\ref{freq_1}).} 
\label{exz_fig}
\end{figure}

Using Eq.\ref{e_thz_kx_c} and \ref{e_thz_x_0} and parameters from Table.\ref{param_list}, we first plot the spatial distribution of terahertz spectra in Fig.\ref{exz_fig}. All calculations do not consider the effects of the crystal boundary on the spectral properties.  In Fig.\ref{exz_fig}(a), the spatial distribution of the terahertz spectrum for a beam with radius $w_0=2$mm and pump bandwidth $\tau_0=500$fs is depicted at the point of maximum efficiency. It can be seen that the spatial distribution of the spectrum is rather uniform in relation to the case of $\tau_0=50$fs in Fig.\ref{exz_fig}(b), where the spectrum and center frequency reduce for larger values of $x$. The greater degree of inhomogeneity for $\tau=50$fs, is consistent with the spatio-temporal snapshots of terahertz transients contrasted in Figs.\ref{e_t_500} and \ref{e_t_50}.In Fig.\ref{exz_fig}(c), the average spectrum corresponding to Fig.\ref{exz_fig} is depicted. In addition, the average spectrum for a larger beam radius of $w_0=1$cm is also shown. Clearly, the average frequency drops for larger beam sizes as is evident from larger terahertz durations anticipated for larger beam radii in Eqs.\ref{tau_1}-\ref{tau_2}. In Fig.\ref{exz_fig}(d), average spectra for $\tau_0=50$fs are shown. Here, the beam radius of 1 cm produces a greater reduction in average frequency due to the greater impact of GVD-AD for shorter pump durations.

As is clearly evident, the strongest influence on central frequency $\Omega_{max}$ is the bandwidth and beam radius $w_0$. While a shorter duration $\tau_0$ produces higher terahertz frequencies of $\Omega_{max} \approx 2/\tau_0$ for $k_T=0$, the effect of GVD-AD causes this value to reduce upon continuous increase of either bandwidth of beam radius $w_0$.

\begin{figure}
\centering
\includegraphics[scale=0.425]{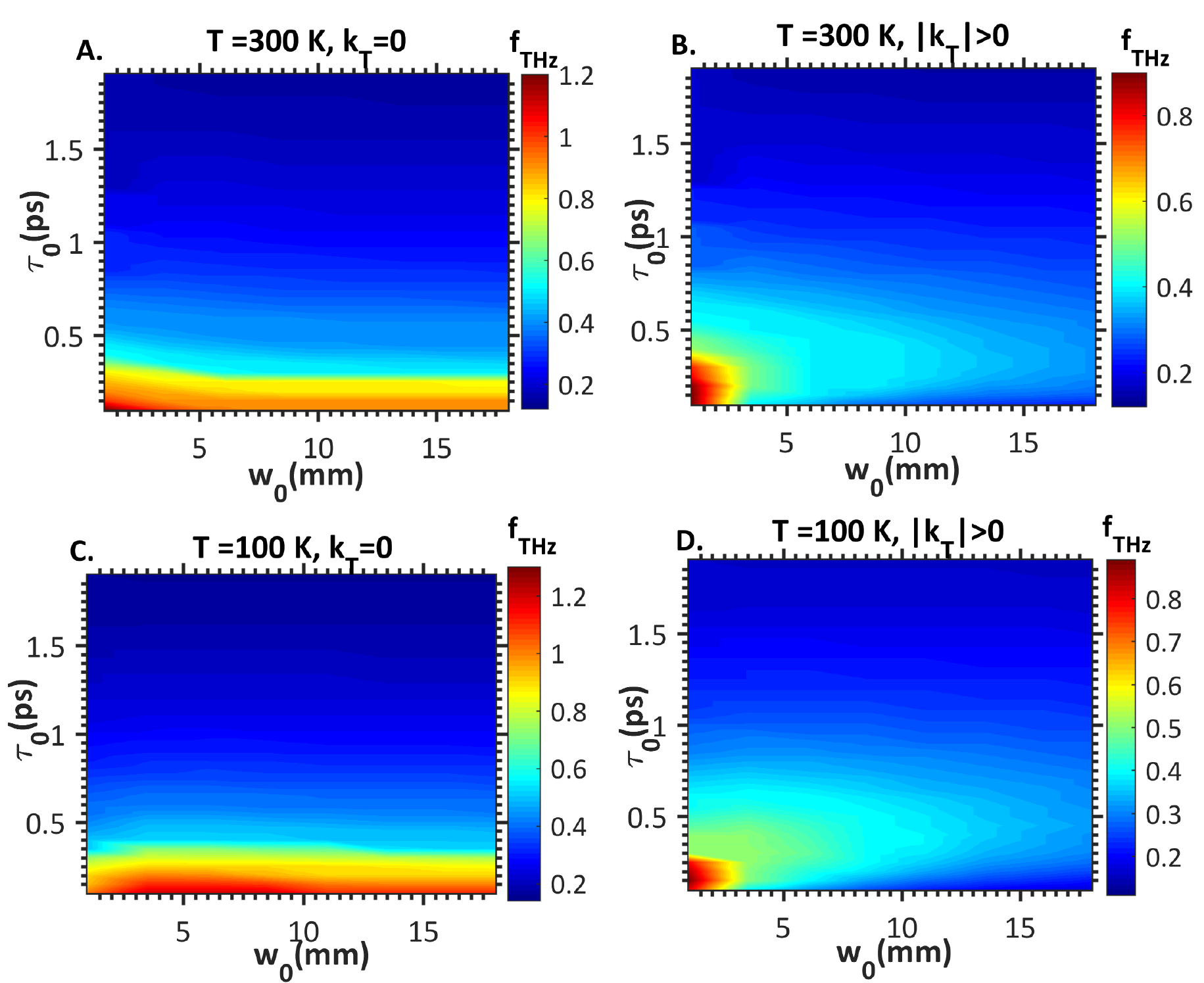}
\caption{Central terahertz frequencies as a function of $\tau_0,w_0$ calculated using Eq.\ref{e_thz_kx_c} in lithium niobate for (a) No GVD-AD, i.e. $k_T=0, T=300 K$ shows a monotonic increase with shorter durations and no $w_0$ dependence, consistent with Eq.\ref{freq_1}.(b) For $k_T=-1.3\times10^{-23}s^2/m$,T=300 K, the reduction of terahertz frequency for large beam sizes is evident.(c) For $k_T=0,T=100$ K, an increase in average frequency compared to (a) due to reduced absorption is seen.(d) For non-zero $k_T, T=100$ K, the frequency is closer to that in (b), indicating relative importance of GVD-AD effects.}
\label{omega_max_fig}
\end{figure}

In Fig.\ref{omega_max_fig}, the central terahertz frequencies for T=300,100K with and without the effects of GVD-AD are depicted using calculations employing Eq.\ref{e_thz_kx_c}.

For T=300 K and $k_T=0$, the maximum frequency occurs at the shortest durations and show no change upon varying beam radius $w_0$. However, for finite values of GVD-AD, the maximum frequency no longer occurs at the shortest duration but at a slightly longer pump duration as is evident in Fig.\ref{omega_max_fig})(b). Furthermore, the dramatic drop in peak frequency with beam radius is evident. In Figs.\ref{omega_max_fig}(c)-(d), the situation for T=100 K is shown. Firstly, in the absence of GVD-AD, the average frequency shows appreciable increase in contrast to the T=300 K case due to reduction in absorption. However, for the case of finite GVD-AD, the reduction in absorption by cryogenic cooling does not appear to influence the peak frequency much, indicating that the frequency is mainly GVD-AD limited.  

\subsection{Conversion Efficiency}
Using Eq.\ref{e_thz_kx_c}, the conversion efficiency $\eta$ is then evaluated for the four cases outlined above, i.e. for T=300,100 K with and without the inclusion of the effects of GVD-AD for various values of $\tau_0,w_0$. The results are plotted in Fig.\ref{eta_fig} below. 

\begin{figure}
\centering
\includegraphics[scale=0.425]{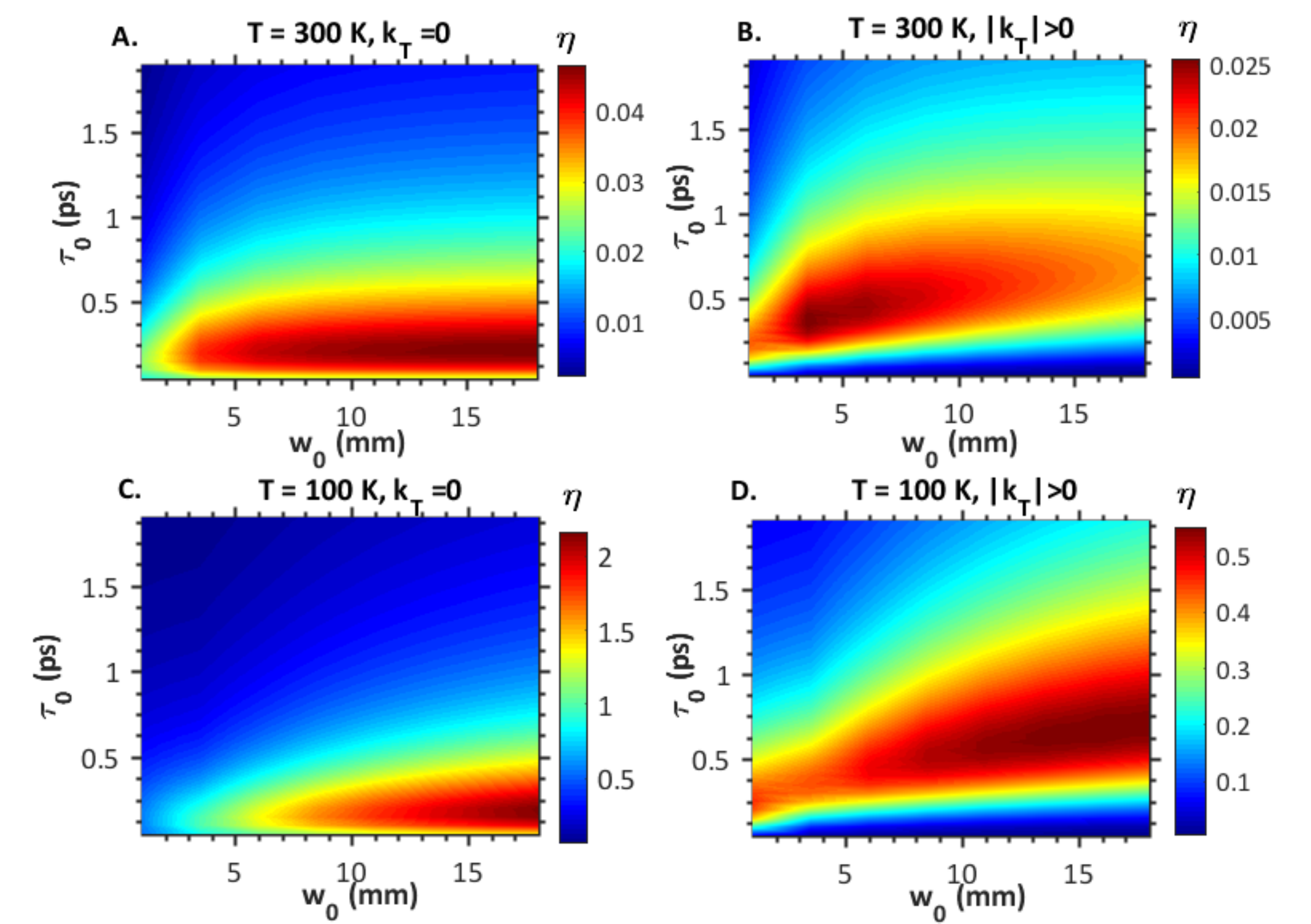}
\caption{Conversion efficiency $\eta$ as a function of $\tau_0,w_0$ based on Eq.\ref{e_thz_kx_c} in lithium niobate.(a)For $k_T=0, T=300$ K, $\eta$ saturates beyond a critical beam radius and shows an optimum value at $\approx 250$fs.(b)For $k_T=-1.3\times10^{-23}s^2/m$ and T=300 K, effects of GVD-AD increase optimal duration as alluded to by Eq.\ref{eta_approx}. A reduction rather than saturation at large $w_0$ is seen (c) For $k_T=0,T=100$K, saturation occurs at much larger $w_0$ due to increased interaction lengths by virtue of reduced absorption.(d) For $|k_T|>0$, peak efficiency occurs at larger $\tau$ compared to (c), consistent with (b)}
\label{eta_fig}
\end{figure}

For T=300 K and $k_T=0$ (or no GVD-AD), conversion efficiency saturates after a certain minimum value of $w_0$, which is consistent with the calculations presented in \citep{bakunov2008,bakunov2011}. An optimal duration exists since for shorter durations, the peak frequency is larger which corresponds to larger absorption coefficients. For the case when GVD-AD effects are included in Fig.\ref{eta_fig}(b), firstly conversion efficiency does not saturate with $w_0$ but instead shows a decline beyond a certain value of $w_0$. Furthermore, the optimal pump durations shift to larger values. In Fig.\ref{eta_fig}(c), the case for T=100 K and $k_T=0$ is shown. Compared to Fig.\ref{eta_fig}(a), the beam radius at which efficiency saturation has increased due to a reduction in absorption, which yields larger interaction lengths. A reduction in the optimal pulse duration accompanying an increase in frequency due to reduced absorption is also evident. In Fig.\ref{eta_fig}(d), the reduction of absorption has increased the interaction length and moved the optimum to larger beam sizes but due to marginal change in frequency, the optimal pulse durations have not changed much. The latter trend is consistent with experiments but the former is not, since at cryogenic temperatures, terahertz generation is limited by cascading effects more adversely compared to absorption.

Similar to the previous section, the influence of GVD-AD on $\eta$ is evident from the closed-form expression for $\eta$ presented for the case of $k_T^2w_0^2/\tau^4\\1$ in Eq.\ref{eta_approx}. It is therefore evident that in the absence of absorption, conversion efficiency should increase with decreasing $\tau$. However, for finite $k_T$, the second term in the square brackets reduces the efficiency for larger beam size and reducing $\tau$, counteracting the $\tau^{-3}$ factor outside the brackets.

If one accounts for a scaling factor of $1/\sqrt{2}$ due to the third spatial dimension $y$ and an additional factor of $1/2$ due to Fresnel losses, then the maximum conversion conversion efficiency in Fig.\ref{eta_fig}(b) for T=300 K is $\approx 0.88\%$ at $\tau \approx 400$fs and $w_0=5$ mm. If the expression for optimal pulse duration from Eq.\ref{tau_opt} is used, then one obtains the value to be $\tau_{opt}=413$ fs, which is very close to the simulated results. This is also consistent with experimental trends \cite{huang2015}. However, the conversion efficiencies from this undepleted model are a bit higher even at room temperatures since cascading effects are not accounted for. These cascading effects in combination with GVD-AD deteriorate phase-matching and reduce the conversion efficiency \cite{ravi2014}. At T=100 K, the conversion efficiency is significantly overestimated. Here, the impact of cascading shall limit performance even further. In summary, undepleted calculations provide good qualitative understanding and fair quantitative predictions at room temperature while overestimating conversion efficiency at cryogenic temperatures.

\subsection{Peak field}

\begin{figure}
\centering
\includegraphics[scale=0.41]{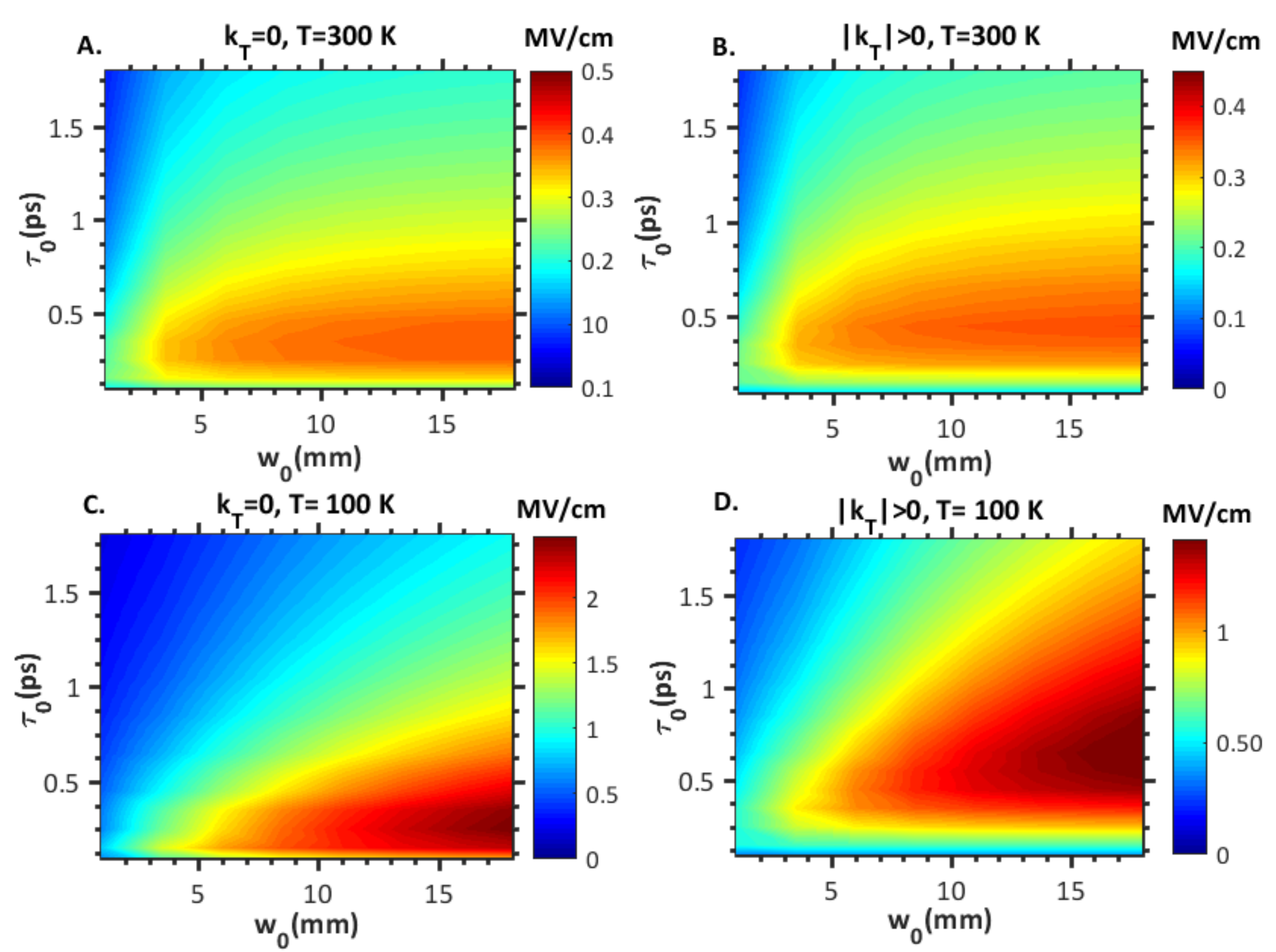}
\caption{Peak electric field obtained from Eq.\ref{e_thz_kx_c}-\ref{e_thz_t_0} for (a)$k_T=0,T=300$K (b)$k_T=-1.3\times10^{-23}s^2/m,T=300 K$(c)$k_T=0,T=100$K and (d)$|k_T|>0,T=100$K. The results are essentially proportional to the product of $\eta$ and $f_{max}$. As a result peak fields are obtained at lower pump durations for $T=100$K whereas they are relatively flat for $T=300 K$. The values are well described by Eq.\ref{e_max} for T=300 K due to satisfaction of constraints in Eq.\ref{constraint_a}-\ref{constraint_2}.}
\label{emax_fig}
\end{figure}

These values are proportional to the ratio of the product of $\eta$ and $\Omega_{max}$ from Figs.\ref{eta_fig} and \ref{omega_max_fig}. For T=300K and $k_T=0$, $\omega_{max}$ is largest for shortest pump durations while $\eta$ is optimized for slightly durations. Therefore, the peak-field strengths show a flatter profile as a function of $\tau$ in Fig.\ref{emax_fig}(a). In Fig.\ref{emax_fig}(b)-(d), a similar effect is observed. Consistent with Fig.\ref{eta_fig}(d), the values of peak-field for T=100 K in Fig.\ref{emax_fig}(d) are inflated due to operating in the undepleted limit.

\section{Conclusion}

An undepleted analysis of terahertz generation by optical rectification of tilted-pulse-fronts was performed. The analysis accounted for the effects of various spatio-temporal distortions imparted to the optical pump beam by the tilted-pulse-front setup such as angular dispersion, spatial and temporal chirp, group velocity dispersion due to material and angular dispersion as well as the radius of curvature of the pump beam. Closed form expressions to evaluate the properties of the generated terahertz radiation were provided in the transverse momentum domain, which are generally valid for all bandwidths and beam sizes. In addition, closed form expressions for terahertz spectra and transients for small $k_T^2w_0^2/\tau^4$ were provided. These closed-form expressions in spatial and temporal domains provide clear qualitative understanding of the physics at work and accurate quantitative predictions at room temperature for high energy terahertz generation systems. The effect of radius of curvature on tilt angle was formally derived and presented along with illustrative calculations relevant to widely employed experimental setups. The overarching point shown via the analysis was the detrimental effect of GVD-AD on beam profiles, terahertz frequency, efficiency and scaling with beam-size. While, this was numerically shown in previous work, here we show this conclusively using analytic formulae in multiple spatial dimensions. The undepleted analysis here is able to provide useful insights into terahertz generation using tilted-pulse-fronts. In the large absorption limit, such as the case for terahertz generation at room temperature in lithium niobate, the model provides reasonable quantitative predictions. In the low absorption limit, the effects of cascading become more important, which would require full numerical simulations.

\section{Appendix}

\begin{table*}[ht]
\begin{tabular}{l c r}
\hline
Symbol& Variable & Units (SI) \\
\hline
$\omega$ & Optical angular frequency & rads/s \\

$\omega_0$ & Central optical  angular frequency & rads/s \\

$k_0$ & Central optical wave number & m $^{-1}$ \\

$k(\omega)$ & Optical wave number  & m$^{-1}$   \\

$\gamma$  & Tilt angle & rads \\

$\Omega$ & Terahertz angular frequency & rads/s\\

$k(\Omega)$ & Terahertz wave number  & m$^{-1}$   \\

$\Delta k$&Phase-mismatch& m$^{-1}$\\

$v_g$& Optical group velocity & m/s\\

$\Delta\omega$ & Displacement from $\omega_0$&rads/s\\

$\tau_0$& Input pump pulse duration & s\\

$\tau$& Effective pump pulse duration & s\\

$\zeta$ & Spatial-chirp & m-s\\

$w_0$ & Input beam radius & m\\

$w$ & Effective beam radius & m\\

$\phi_0$ & Input group delay dispersion & s$^2$m$^{-1}$\\

$\phi$ & Effective group delay dispersion & s$^2$m$^{-1}$\\

$R_0$ & Input Radius of curvature & m\\

$R$ & Effective Radius of curvature & m\\

$k_m$& Group velocity dispersion due to material dispersion&  s$^2$m$^{-1}$\\

$\beta_0$ & Input angular dispersion & rads-s\\

$\beta$ & Effective angular dispersion & rads-s\\

$k_T$& Group velocity dispersion due to angular dispersion&  s$^2$m$^{-1}$\\

$x_0$& Central beam position & m \\   
\hline
\end{tabular}
\caption{\label{var_list} List of variables} 
\end{table*}

\subsection{Fourier Transform relations}
We begin by decomposing the real scalar electric in space ($r$) and time ($t$) as $E(r,t) = [f(r,t)e^{-j\omega_{ref}t} + f^{*}(r,t)e^{j\omega_{ref}t}]/2$. The corresponding Fourier transform of the real field is given by $E(r,\omega) = [f(r,\omega-\omega_{ref})+f^{*}(r,\omega+\omega_{ref})]/2$. The reference angular frequency $\omega_{ref}$ is used here only to delineate the fact that the spectrum of the real field comprises of positive and negative frequency components, symmetric and conjugate about $\omega=0$. It may be dropped henceforth. Furthermore, the dependency on $r$ shall be assumed to be implicit and will also be dropped. From the above it is clear that $f(\omega)$ is a Fourier transform of $f(t)$. The conventions in Eq.\ref{FT_t} are assumed for Fourier transforms between temporal and spectral domains.  

\begin{gather}
f(\omega) = \mathfrak{F}_t\lbrace f(t)\rbrace=\frac{1}{2\pi}\int_{-\infty}^{\infty}f(t)e^{-j\omega t}dt \nonumber\\
f(t)  = \mathfrak{F}^{-1}_t \lbrace f(\omega)\rbrace= \int_{-\infty}^{\infty}f(\omega)e^{j\omega t}d\omega \label{FT_t}
\end{gather}

Similarly, for spatial Fourier transforms in the transverse dimension $x$ are provided by Eq.\ref{FT_x}.

\begin{gather}
g(k_x) = \mathfrak{F}_x\lbrace g(x)\rbrace= \frac{1}{2\pi}\int_{-\infty}^{\infty}g(x)e^{jk_x.x}dx\nonumber\\
g(x) = \mathfrak{F}^{-1}_x\lbrace g(k_x) \rbrace=\int_{-\infty}^{\infty}g(k_x)e^{-jk_x x}dk_x\label{FT_x}
\end{gather}

The energy in each space/time is equal to that in transverse momentum/angular frequency according to Parseval's theorem in Eq.\ref{parseval}.

\begin{gather}
\int_{-\infty}^{\infty}|f(x)|^2dx = 2\pi\int_{-\infty}^{\infty}|f(k_x)|^2dk_x \label{parseval}
\end{gather}

In evaluating Eq.\ref{wav_eq1}, it would suffice to book-keep for terahertz angular frequencies $\Omega>0$ for $E_{THz}(\Omega,x,z)$, which corresponds to the $f(\omega-\omega_{ref})$ term above. Therefore, in evaluating the real electric field $E_{THz}(t,x,z)$, we adopt Eq.\ref{e_thz_kx_c}. In practice, since $E_{THz}(-\Omega,x,z)=E^{*}_{THz}(\Omega,x,z)$ readily, one may evaluate $E_{THz}(t,x,z)= \mathfrak{F}^{-1}_t E_{THz}(\Omega,x,z)/2$ (for positive and negative values of $\Omega$).

The energy per unit length of real electromagnetic field vectors $\vec{\textbf{E}}$(electric) and $\vec{\textbf{H}}$ (magnetic) passing through an area of cross-section with unit vector $dxdy~\hat{z}$ and refractive index $n$ is given by $\int_{-\infty}^{\infty}\int_{-\infty}^{\infty}\vec{\textbf{E}}(r,t)\times\vec{\textbf{H}}(r,t).\hat{z}dxdt= c\varepsilon_0 n\int_{-\infty}^{\infty}\int_{-\infty}^{\infty}|E(r,t)|^2dxdt$. In terms of the phasor quantity $f(t)$, this translates to $\frac{1}{2}c\varepsilon_0 n\int_{-\infty}^{\infty}\int_{-\infty}^{\infty}|f(r,t)|^2dxdt$.

\subsection{Notes for the derivation of Eq.\ref{e_thz_kx_a}}

For integrals of the following form :
\begin{gather}
\int_0^z P_{THz}(\Omega,k_x,z')e^{j\Delta kz' +\frac{\alpha}{2}z'}dz'\nonumber\\
=\frac{P_{THz}(\Omega,k_x,z)e^{j\Delta kz +\frac{\alpha}{2}z}}{\alpha/2+j\Delta k}-\frac{P_{THz}'(\Omega,k_x,z)e^{j\Delta kz +\frac{\alpha}{2}z}}{(\alpha/2+j\Delta k)^2}+\nonumber\\
\frac{P_{THz}"(\Omega,k_x,z)e^{j\Delta kz +\frac{\alpha}{2}z}}{(\alpha/2+j\Delta k)^3}..
\end{gather}

Naturally, for relatively large $\alpha$ and $P'_{THz}(\Omega,k_x,z)/P_{THz}(\Omega,k_x,z)$, only the first term in the expansion shall be significant. The variation along space of the Polarization term due to material is typically small while absorption coefficient values even at cryogenic temperatures for lithium niobate $126$/m are quite large to result in the validity of this approximation. 

\section{Acknowledgments}

This work was supported under by the Air Force Office of Scientific Research under grant AFOSR - A9550-12-1-0499
the European Research Council under the European Union's Seventh Framework Programme (FP/2007-2013) / ERC Grant Agreement n. 609920 and the Center for Free-Electron Laser Science at DESY.

\bibliography{sample}

\end{document}